\documentclass[aps,prd,onecolumn,groupedaddress,amssymb,showpacs,nofootinbib,epsfig,pdffig]{revtex4}
\usepackage{graphicx}
\usepackage{bm}
\usepackage{dcolumn}
\usepackage{amsmath}
\usepackage{amssymb}
\usepackage{epsfig}
\usepackage{color}
\usepackage[colorlinks,linkcolor=blue,citecolor=blue,urlcolor=blue]{hyperref}

\begin{document}

\title{Initial conditions of pre-inflation with Hilltop potential in loop quantum cosmology}
\author{M. Shahalam$^{1,3}$ \footnote{E-mail address: mohdshahamu@gmail.com}}
\author{Kuralay  Yesmakhanova$^{2,3}$ \footnote{ E-mail address:kryesmakhanova@gmail.com}}
\author{Zhanar Umurzakhova$^{2,3}$ \footnote{ E-mail address:zumurzakhova@gmail.com}}
\affiliation{$^{1}$Department of Physics, Aligarh Muslim University, Aligarh 202002, India\\
$^2$Eurasian National University, Nur-Sultan 010008, Kazakhstan\\
$^3$Ratbay Myrzakulov Eurasian International Centre for Theoretical Physics, Nur-Sultan 010009, Kazakhstan\\
}

\date{\today}

\begin{abstract}
In this paper, we investigate the dynamics of pre-inflation with Hilltop potential in the framework of loop quantum cosmology. The initial conditions of inflaton field at the quantum bounce is categorized into two classes, first one is dominated by kinetic energy and second one by potential energy. In both cases, the physically viable initial values of inflaton field at the bounce are obtained numerically that generate the desired slow-roll inflation and also sufficient number of $e$-folds. To be consistent with observations at least 60 $e$-folds are required. In case of kinetic energy dominated (KED) initial conditions of inflaton field at the bounce, the numerical evolution of the background prior to preheating is divided into three different regions: {\it bouncing, transition and slow-roll inflation} whereas bouncing and transition phases disappear in the potential energy dominated (PED) case but still slow-roll inflation is achieved. This is true in case of KED (except subset) and PED initial conditions for $p=4$ and $v=1 M_{Pl}$ of Hilltop potential. However for other cases, slow-roll inflation can not be obtained. Moreover, we study the phase space analysis for Hilltop potential and discuss the phase space trajectories under the chosen parameters. 

\end{abstract}
\maketitle

\section{Introduction}
\label{sec:intro}
The cosmic inflation is an accelerated expansion of space in the early universe. The phase of accelerated expansion of universe is driven by a scalar field, known as inflaton. Cosmic inflation explains the initial conditions of big bang theory i.e. it  resolves various problems in the standard model of cosmology such as the  flatness and horizon problems. Inflation describes the origin of inhomogeneities in the cosmic microwave background and the structure formation of the universe \cite{guth1981}. A large variety of inflationary models have been suggested in the literature, namely,  Starobinsky, $\alpha-$attractor and the chaotic inflation etc. \cite{staro1980,staro4,alpha,alpha1,alpha2,GL}. These models are consistent with the current observations. However, future observations may be narrow down some of the viable models. According to Planck 2018 observations, quadratic potential is completely disfavored whereas Hilltop, $\alpha-$attractor and Starobinsky potentials are consistent with data \cite{Planck2018}. In this paper, we shall study the dynamical behavior of pre-inflationary universe with Hilltop potential in the framework of loop quantum cosmology (LQC), and find out the initial conditions of pre-inflation numerically, at the quantum bounce. Further, we shall show that the initial conditions of inflaton field at the bounce are compatible with slow-roll inflation or not. In classical theory of general relativity (GR), all scalar field models of inflation experience the big bang singularity which is inevitable \cite{borde1994,borde2003}. To this effect,  it is  hard to know when and how to impose the initial conditions. To be compatible with observations, the number of $e$-folds during inflation should be at least 60. Meanwhile, in some cases, the number of $e$-folds is more than 70 \cite{martin2014}. However, in such kind of models the size of present universe is smaller than the Planck at the starting of inflation. As a result, the semi-classical treatments are questionable during inflation which is so-called  trans-Planckian problem \cite{martin2001,berger2013}. 

Above issues can be addressed in the context of LQC that gives viable explanation of inflation and pre-inflation, simultaneously. In LQC, the big bang singularity is replaced by a non-singular quantum bounce  \cite{agullo2013a,agullo2013b,agullo2015,ashtekar2011,ashtekar2015,barrau2016}. We shall examine the dynamics of pre-inflation with Hilltop potential in the framework of LQC, and explore whether following the bounce a desired slow-roll inflation is obtained or not \cite{ashtekar2010,psingh2006,zhang2007,chen2015,bolliet2015,schander2016,bolliet2016,Bonga2016,Mielczareka}.
In this work, we are mostly concerned with the numerical evolution of the background.
In particular, we shall show that the numerical evolution of the universe before preheating can be divided universally into three different phases: {\em bouncing, transition and slow-roll inflation} in case of the kinetic energy dominated (KED) initial conditions whereas bouncing and transition phases disappear in potential energy dominated (PED) case. 

The paper is organized as follows. In Sec. \ref{sec:EOM}, we discuss the background equations with a spatially flat Friedmann-Lemaitre-Robertson-Walker (FLRW) universe in the context of LQC. Sec. \ref{sec:pot} is devoted to the Hilltop inflation where we study the spectral index $n_s$ and tensor to scalar ratio $r$ for said model. In addition, we investigate the background evolution for Hilltop potential, and conclude that whether the desired slow-roll inflation with at least 60 $e$-folds is achieved or not. The phase space trajectories are demonstrated in Sec. \ref{sec:port}, and the results are summarized in Sec. \ref{sec:conc}.

\section{Equations of Motion in LQC}
\label{sec:EOM}
In this section, we study the evolution equations with a spatially flat FLRW background in the framework of LQC. The quantum corrected Friedmann equation and the Klein-Gordon equation for a single scalar field are written as \cite{ashtekar2006}
\begin{eqnarray}
H^2=\frac{8 \pi}{3 m_{Pl}^2}~\rho \Big{(}1-\frac{\rho}{\rho_c}\Big{)}, 
\label{eq:Hub}
\end{eqnarray}
\begin{eqnarray}
\ddot{\phi}+3H \dot{\phi}+ \frac{dV(\phi)}{d\phi}=0.
\label{eq:ddphi}
\end{eqnarray}
where $H=\dot{a}/a$ designates the Hubble parameter, the dot denotes a derivative with respect to the cosmic time $t$ and $m_{Pl}$ is the Planck mass. The energy density of the inflaton field is $\rho=\dot{\phi}^2/2+V(\phi)$,  and $V(\phi)$ represents the potential of the field. The correction term $-\rho^2/\rho_c$ comes due to the quantum geometric effects. In classical limit, the critical energy density $\rho_c \rightarrow \infty $, and one can recover the original Friedmann equation as given by GR. The value of critical energy density is provided as 
$\rho_c \simeq 0.41 m_{Pl}^4$ \cite{ashtekar2006} which is maximum value of energy density in LQC. It is remarkable to see that the correction term has negative sign that permits the quantum bounce without the violation of energy condition unlike GR. Following equation (\ref{eq:Hub}), we conclude that $H=0$ at $\rho=\rho_c$ this means that bounce occurs when $\rho$ reaches $\rho_c$. In the literature, the extensive work has been studied with the bouncing phase by using the background equations of motion. One of the vital result is that one can obtain the desired slow-roll inflation \cite{psingh2006,Mielczarek,zhang2007,chen2015,Tao2017a,Tao2017b,alam2017,alam2018,alam1,alam2,alam3}. In continuation to this, we shall examine ``bounce and slow-roll inflation" with the Hilltop potential (see Sec. \ref{sec:pot}).

Before moving to the specific potential (Hilltop), let us first explore the background equations for a general potential $V(\phi)$. We  solve  Eqs.(\ref{eq:Hub}) and (\ref{eq:ddphi}) numerically with the initial values of $a(t)$, $\phi(t)$ and $\dot{\phi}(t)$ given at a specific time. One of the possibility of time is at the bounce $(t=t_B)$, for which we have
\begin{eqnarray}
\rho &=& \frac{1}{2}\dot{\phi}^2(t_B)+V(\phi(t_B))=\rho_c, \nonumber\\
\dot{a}(t_B)&=&0, 
\label{eq:bounce}
\end{eqnarray}
from which we get the inflaton velocity as
\begin{eqnarray}
\dot{\phi}(t_B) &=& \pm \sqrt{2 \Big{(} \rho_c - V(\phi(t_B)) \Big{)}}.
\label{eq:bounce2}
\end{eqnarray}
In this paper, we shall use positive sign of equation (\ref{eq:bounce2}) to solve the background equations. However, one can also work with negative inflaton velocity (NIV) to get similar results. Therefore, we restrict ourselves to choose the positive inflaton velocity (PIV) in sections \ref{sec:EOM} and \ref{sec:pot}. Though, both positive and negative signs of inflaton velocity will be used in Sec. \ref{sec:port} to show the phase space trajectories in whole phase space. Without loss of the generality, we can always pick 
\begin{eqnarray}
a(t_B) &=& 1.
\label{eq:bounce3}
\end{eqnarray}
For the sake of simplicity, we shall denote $\phi(t_B)$ and $\dot{\phi}(t_B)$ by $\phi_B$ and $\dot{\phi}_B$ in the subsequent sections. It is clear from  Eq.(\ref{eq:bounce2}), the initial values will be given by $\phi_B$ only for a given potential. Second, let us define some important quantities that are essential for this paper such as the equation of state $w(\phi)$, the slow-roll parameter $\epsilon_H$ and the number of $e$-folds $N_{inf}$, and are given as  \cite{alam2017,Tao2017a,Tao2017b}
\begin{eqnarray}
\label{eq:w}
w(\phi) &=& \frac{\dot{\phi}^2/2-V(\phi)}{\dot{\phi}^2/2+V(\phi)},\\
\epsilon_H &=& - \frac{\dot{H}}{H^2}.
\label{eq:epsilon}
\end{eqnarray}
During the slow-roll inflation, $w(\phi)\simeq-1$ and $\epsilon_H \ll 1$. Looking at eq. (\ref{eq:w}), we notice
\begin{equation}
 w(\phi) \Big{\vert}_{\phi=\phi_B}
= \begin{cases}  > 0 ~\text{for KE} > \text{PE}, \\
  = 0 ~\text{for KE}=\text{PE}, \\
 < 0 ~\text{for KE} < \text{PE}. \end{cases}
\label{eq:wb}
\end{equation}
\begin{eqnarray}
N_{inf} = ln \Big{(} \frac{a_{end}}{a_i} \Big{)} =  \int_{t_i}^{t_{end}} H(t) dt 
 = \int_{\phi_i}^{\phi_{end}} \frac{H}{\dot{\phi}} d\phi \simeq \int_{\phi_{end}}^{\phi_i} \frac{V}{V'{(\phi)}} d\phi.
\label{eq:Ninf}
\end{eqnarray}
where KE (PE) stands for kinetic energy (potential energy). The $a_i$ is the expansion factor at the onset of inflation and $a_{end}$ when inflation ends,  i.e. $\ddot{a}(t_i) \gtrsim 0$ and  $w(\phi_{end})=-1/3$. Moreover, the analytical expression of the expansion factor is given as \cite{alam2018}
\begin{eqnarray}
a(t) &=& a_B \left( 1+ \delta \frac{t^2}{t_{Pl}^2} \right)^{1/6}.
\label{eq:a}
\end{eqnarray}
where $a_B=a(t_B) \equiv 1 $, $t_{Pl}$ is the Planck time, and $\delta = {24 \pi \rho_c}/{m_{Pl}^{4}}$ represents a dimensionless parameter. Equation (\ref{eq:a}) is only valid in the bouncing regime of the KED bounce. In the following section, we shall examine the initial conditions of pre-inflation with Hilltop potential by choosing the PIV ($\dot{\phi_B}>0$) at the quantum bounce.
\begin{figure*}[tbp]
\begin{center}
\begin{tabular}{cc}
{\includegraphics[width=2.5in,height=2in,angle=0]{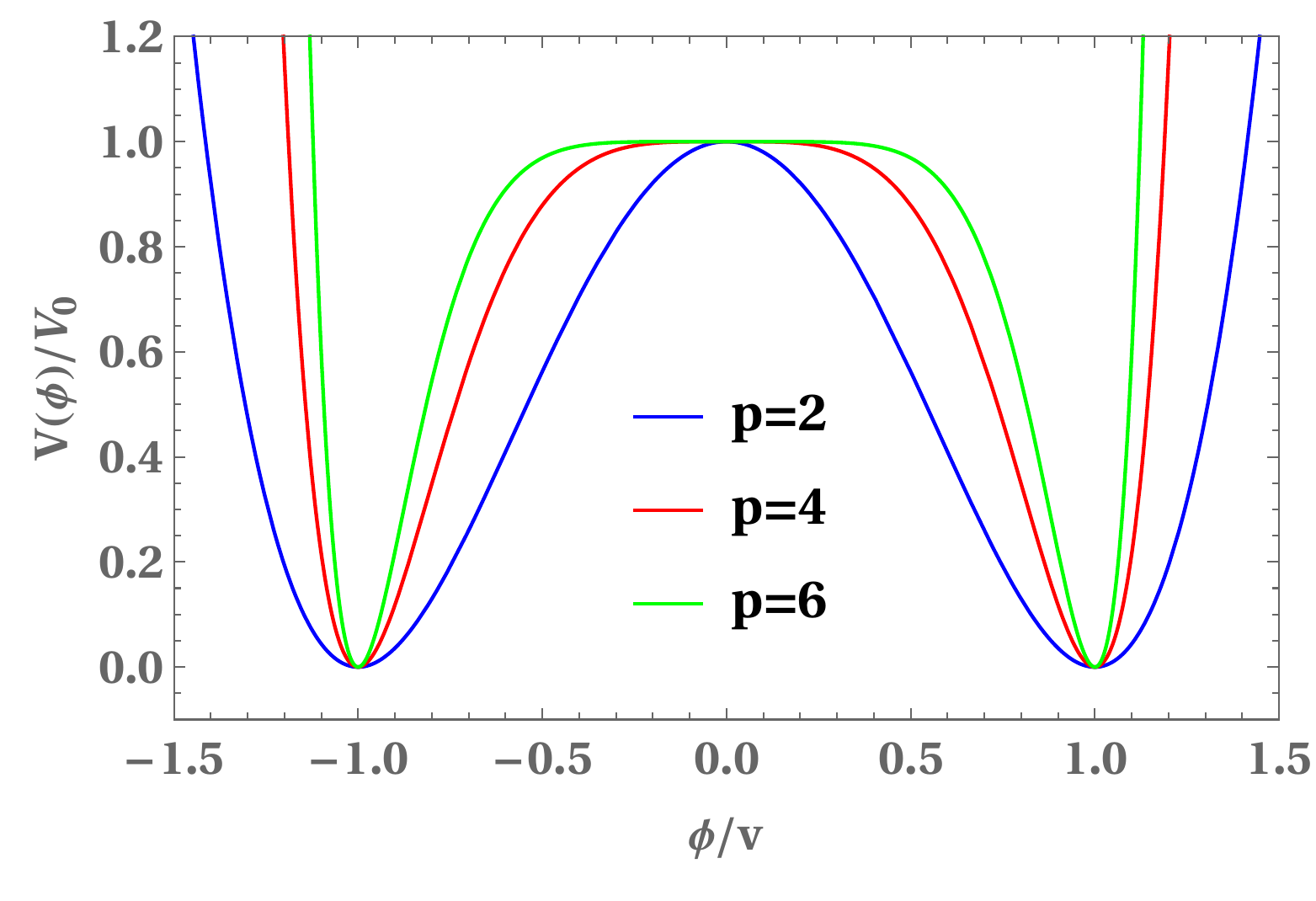}} &
{\includegraphics[width=2.5in,height=2in,angle=0]{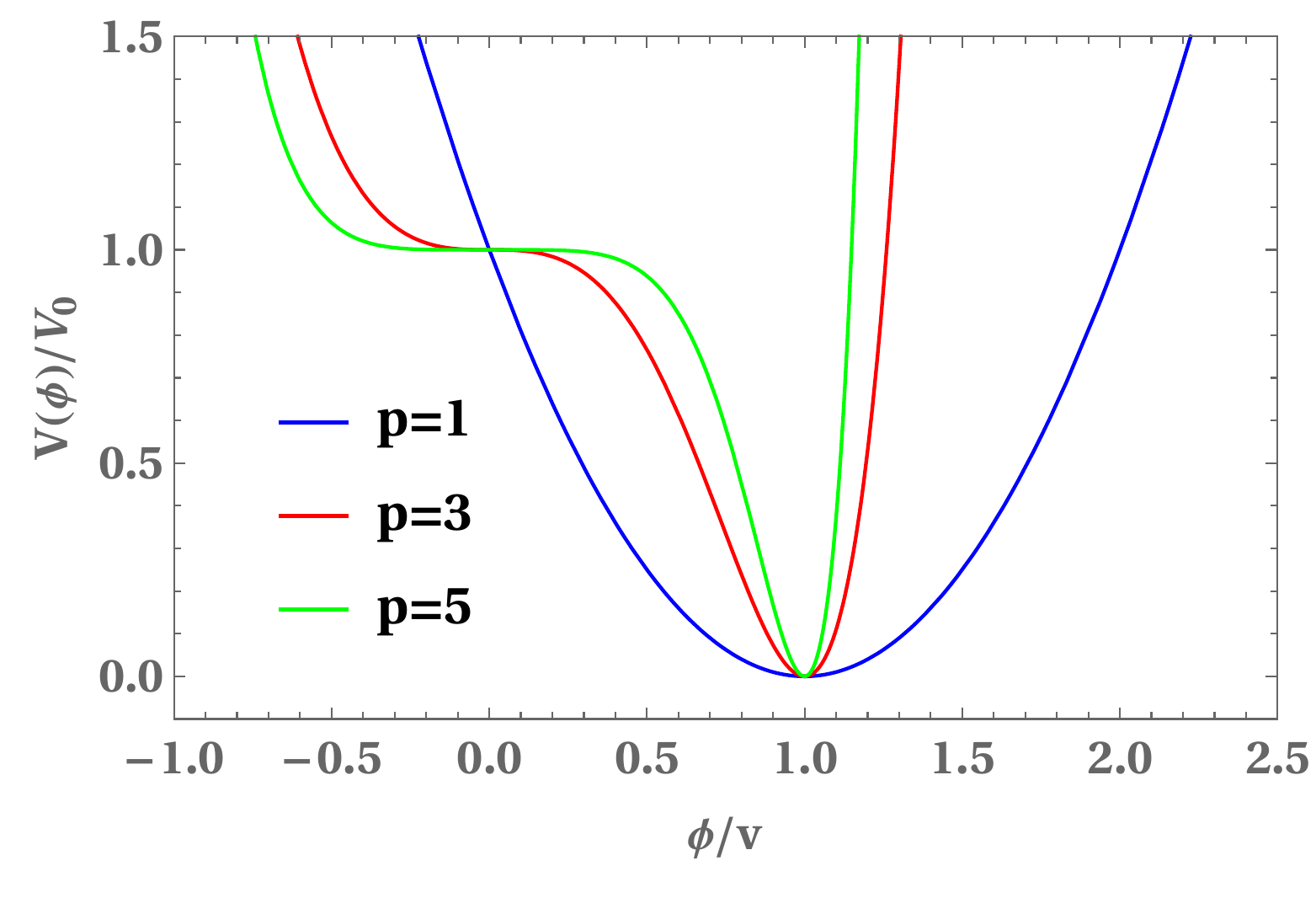}} 
\end{tabular}
\end{center}
\caption{ The figure is schematically presented for Hilltop potential (\ref{eq:pot}) with even (left panel) and odd (right panel) $p's$. For even $p$, it has maximum value at $\phi/v=0$ whereas minimum value at $\phi/v=\pm 1$. For $\phi/v=\pm 1$, the potential is bounded below by zero while unbounded for  $\phi/v \rightarrow \pm \infty $. For odd values of $p$, the potential has similar properties as in case of even $p$ except one minimum at $\phi/v= + 1$. }
\label{fig:pot}
\end{figure*}

\begin{figure*}[tbp]
\begin{center}
\begin{tabular}{cc}
{\includegraphics[width=2.5in,height=2in,angle=0]{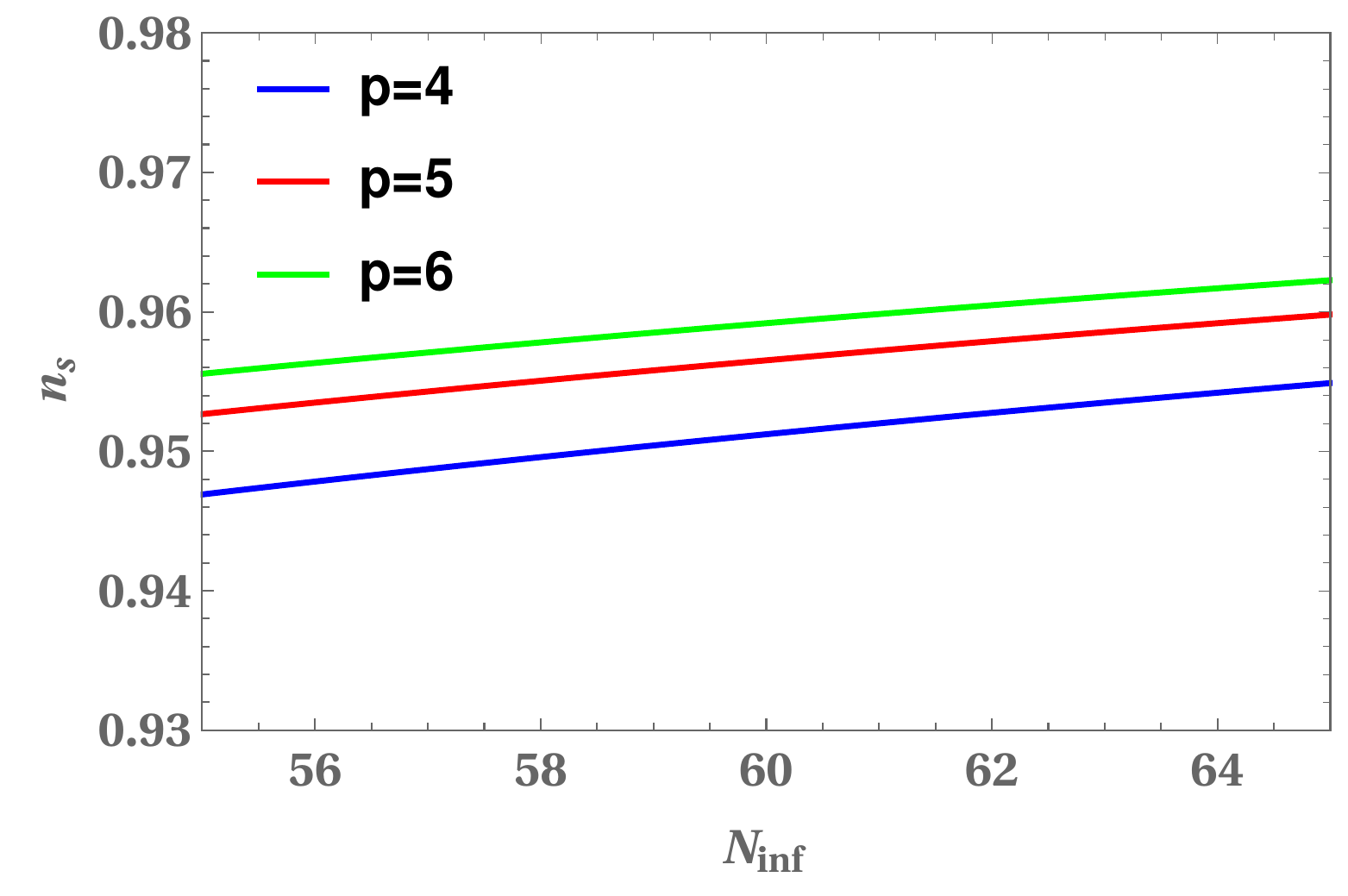}} &
{\includegraphics[width=2.5in,height=2in,angle=0]{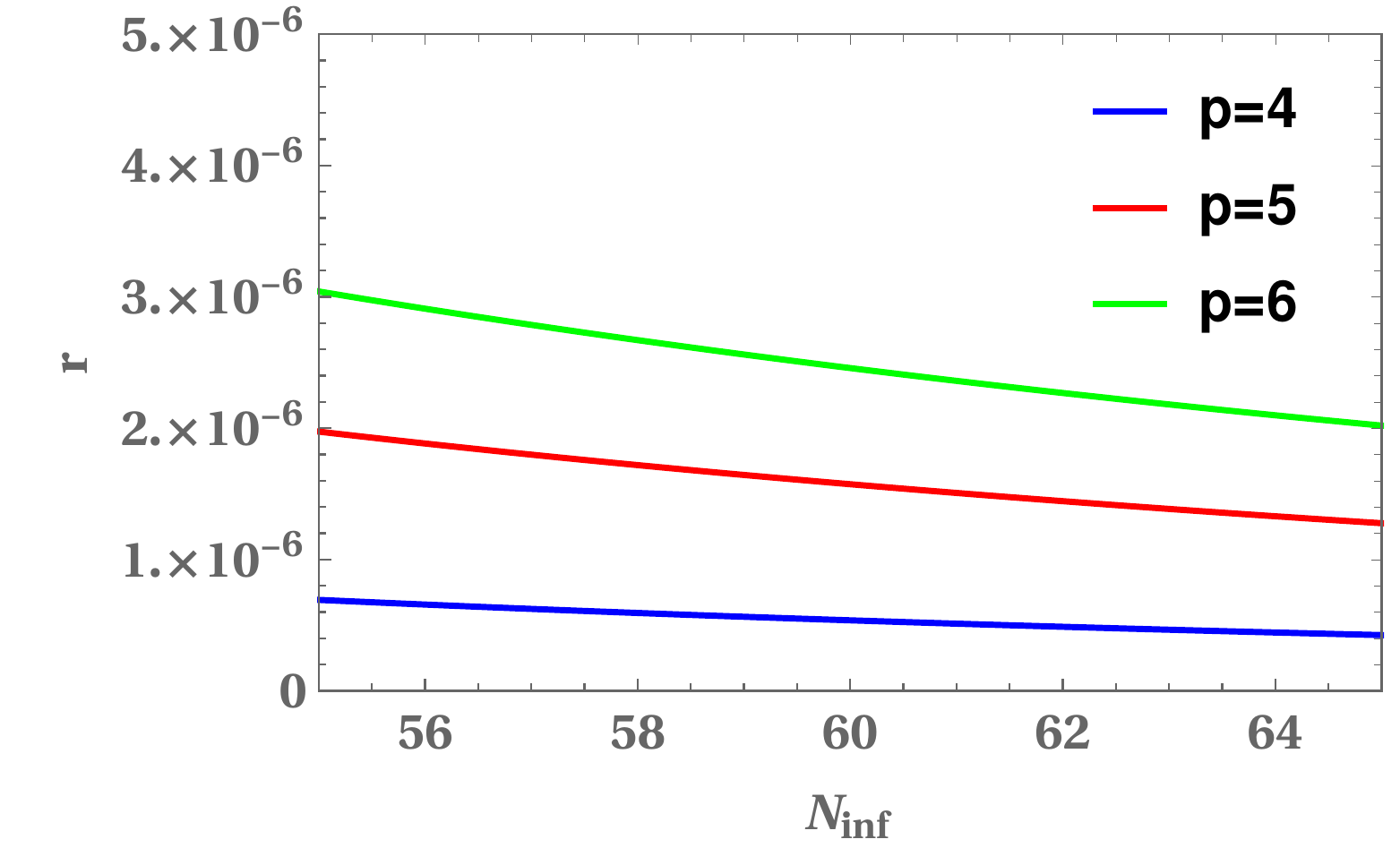}} 
\end{tabular}
\end{center}
\caption{ This figure shows the evolution of spectral index $n_s$ and tensor to scalar ratio $r$ versus number of $e$-folds ($N_{inf}$) for different values of $p$. }
\label{fig:nsr}
\end{figure*}

\section{Hilltop Potential}
\label{sec:pot}
The Hilltop inflation model is supported by the CMB observation, and has compelling connection to particle physics where a phase transition at high energies occurs \cite{pp1,pp2}. Consider a simple realization of Hilltop Potential as \cite{pot}
\begin{equation}
V(\phi)=V_0 \left(  1- \frac{\phi^p}{v^p}\right)^2.
\label{eq:pot}
\end{equation}
where $v \leq M_{Pl} $ and $V_0$ is constant that is constrained by CMB observations. The illustration of Hilltop potential (\ref{eq:pot}) is shown in Fig. \ref{fig:pot} for even and odd values of $p$. Such potentials are highly asymmetric around the minimum. From left panel (even $p$) of Fig. \ref{fig:pot}, one can see that the potential has a local maximum at $\phi=0$ whereas it has global minima at $\phi/v= \pm 1$. For odd $p$, it has one minimum at $\phi/v= + 1$, see right panel of Fig. \ref{fig:pot}. In this paper, we shall work with $\phi>0$. The potential has inflection point towards the plateau for $\phi<v$ and exhibits steeper behavior than quadratic for $\phi>v$. The initial conditions of inflaton field generate the inflation when the potential at $\phi=0$ moves slowly towards the minimum of the potential ($\phi=+v$), and inflation ends when the potential gains large curvature.

Inflation ends at the following value of the inflation field
\begin{equation}
\phi_e= \left( \frac{v^p}{2p(p-1)M_{Pl}^2} \right)^{\frac{1}{p-2}} \qquad \text{when} \qquad \eta(\phi_e)=M_{Pl}^2 \frac{V''(\phi)}{V(\phi)} \simeq -1.
\label{eq:phi_e}
\end{equation}
The scalar field value at the Horizon crossing $\phi=\phi_*$ can be found as
\begin{equation}
\phi_*= \left(\frac{2p M_{Pl}^2}{v^p} \Big{(} (p-2)N_{inf}+(p-1) \Big{)}  \right)^{\frac{1}{2-p}}.
\label{eq:phi_*}
\end{equation}
where $N_{inf}$ is the number of $e$-folds between horizon exist and the end of inflation. The predictions for the spectral index $n_s$ and tensor to scalar ratio $r$ are given by
\begin{eqnarray}
\label{eq:nx}
n_s & \simeq & 1-6 \epsilon(\phi_*)+ 2 \eta(\phi_*) \simeq 1 + 2 \eta(\phi_*),\\
r & \simeq & 16 \epsilon(\phi_*),\\
\label{eq:r}
\text{where} \qquad
\epsilon(\phi_*)&=& \frac{M_{Pl}^2}{2}  \left( \frac{V'(\phi)}{V(\phi)}\right)^2\Big{\vert}_{\phi=\phi_*},\\
\label{eq:ep_*}
\eta(\phi_*)&=&  \frac{1-p}{(p-2)N_{inf}+(p-1)}.
\label{eq:eta_*}
\end{eqnarray}
The spectral index $n_s$ and tensor to scalar ratio $r$ vs $N_{inf}$ for different values of $p$ are displayed in Fig. \ref{fig:nsr}. According to Planck 2018 results, the bound on $n_s$ and $r$ are $n_s=0.9649 \pm 0.0042$ and $r<0.11$ \cite{Planck2018}. Let us see how to fix the value of $V_0$ from the observation of CMB by using the observed value of scalar amplitude $A_s \simeq 2.09 \times 10^{-9}$ \cite{Planck2018}.
\begin{eqnarray}
V_0= 24 \pi^2 \epsilon(\phi_*) A_s M_{Pl}^4 \simeq \frac{48 p^2 \pi^2 A_s M_{Pl}^{2p+4}}{v^{2p}} \left( \frac{v^p}{2p M_{Pl}^{p} \Big{(} (p-2)N_{inf}+(p-1) \Big{)}} \right)^{\frac{2(p-1)}{p-2}}. 
\label{eq:V0}
\end{eqnarray}
In this paper, we shall work with $p=4$ and 5 with $v=0.1 M_{Pl}$ and $1 M_{Pl}$, respectively. Therefore, the corresponding values of $V_0$ can be found by equation (\ref{eq:V0}). The units of equation (\ref{eq:V0}) are given in reduced Planck mass $M_{Pl}$, and the Friedmann equation (\ref{eq:Hub}) is written in  Planck mass $m_{Pl}$. Therefore, we shall convert equation (\ref{eq:V0}) in Planck mass through the relation $M_{Pl}=m_{Pl}/ \sqrt{8 \pi}$, and work in the $m_{Pl}$ unit throughout the paper. The model parameters that will be used in the paper are given as
\begin{eqnarray}
p=4, \qquad v&=&0.1  M_{Pl}=0.019 m_{Pl}\qquad \text{and} \qquad V_0=2.632 \times 10^{-21} m_{Pl}^4,\nonumber\\
 v&=&1.0  M_{Pl} =0.19~ m_{Pl}\qquad \text{and} \qquad V_0=2.632 \times 10^{-17} m_{Pl}^4,\nonumber\\
p=5, \qquad v&=&0.1  M_{Pl} =0.019 m_{Pl} \qquad \text{and} \qquad V_0=3.577 \times 10^{-20} m_{Pl}^4,\nonumber\\
 v&=&1.0  M_{Pl} =0.19~ m_{Pl}\qquad \text{and} \qquad V_0=7.708 \times 10^{-17} m_{Pl}^4.
\label{eq:pvV0}
\end{eqnarray}

In Table \ref{tab-v}, we display the range of $v$ (depends on $\phi_B$)  having slow-roll (SR) and non slow-roll (NSR) inflation for potential (\ref{eq:pot}) with $p=$ 4 and 5. In the continuation, we choose $v=0.1 M_{Pl}$ and $1 M_{Pl}$ in each case of $p$, and draw the figures for various initial conditions of inflaton field. Moreover, the corresponding ranges of inflaton field at the bounce for SR and NSR inflation is shown in Table \ref{tab-phiB}.

\begin{figure*}[tbp]
\begin{center}
\begin{tabular}{ccc}
{\includegraphics[width=2.1in,height=1.65in,angle=0]{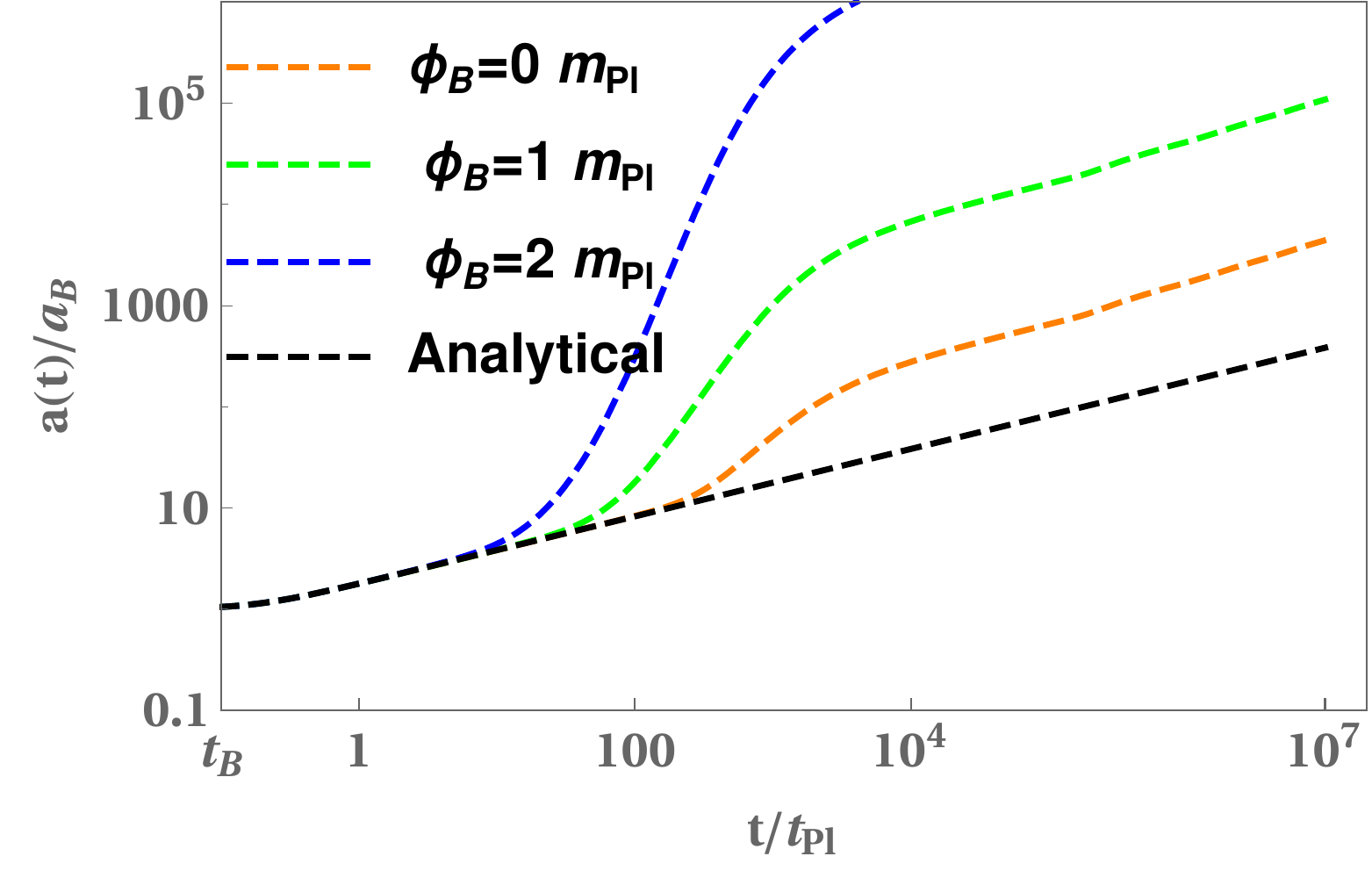}} &
{\includegraphics[width=2.1in,height=1.6in,angle=0]{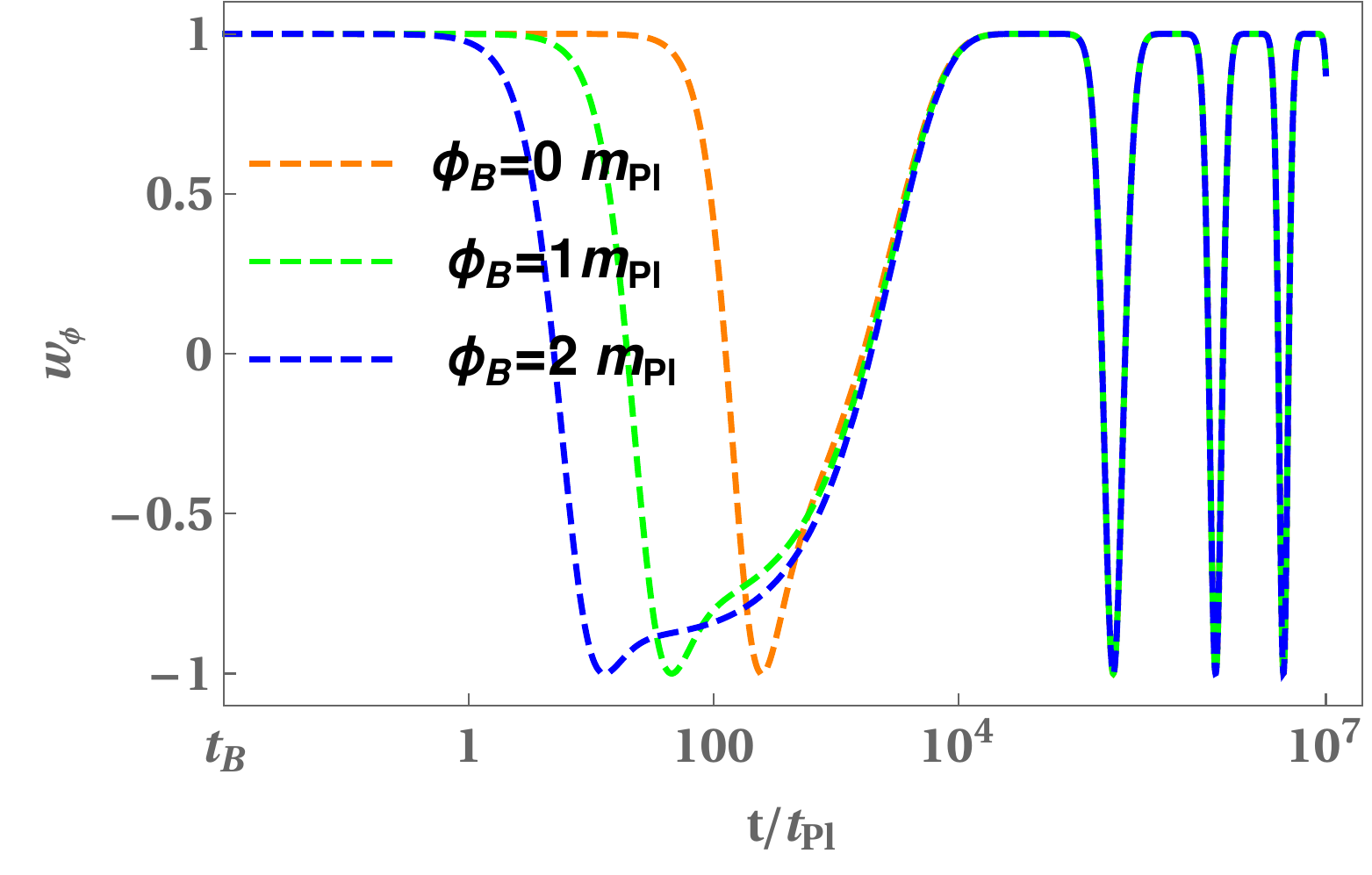}} &
{\includegraphics[width=2.0in,height=1.6in,angle=0]{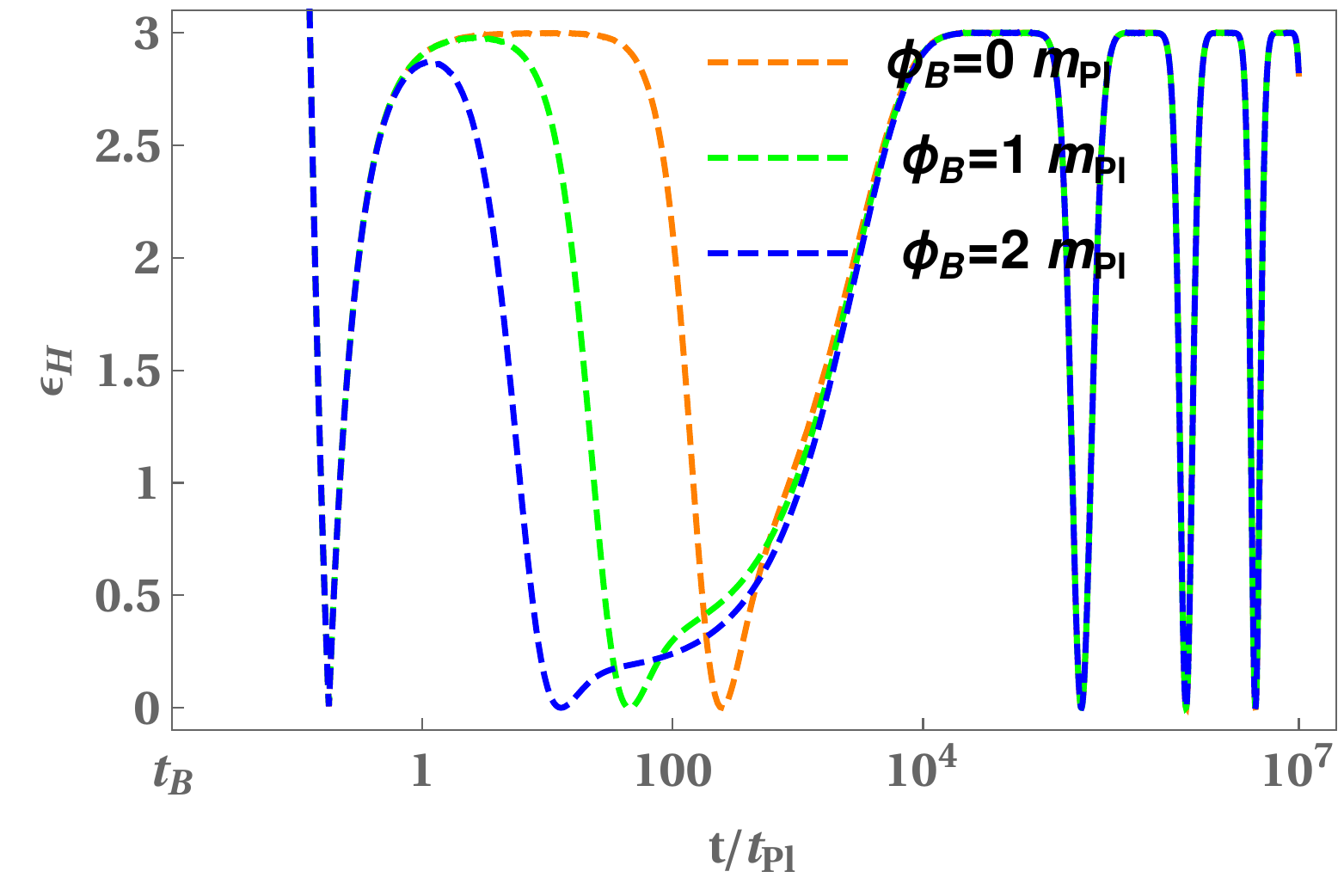}}
\\
{\includegraphics[width=2.1in,height=1.6in,angle=0]{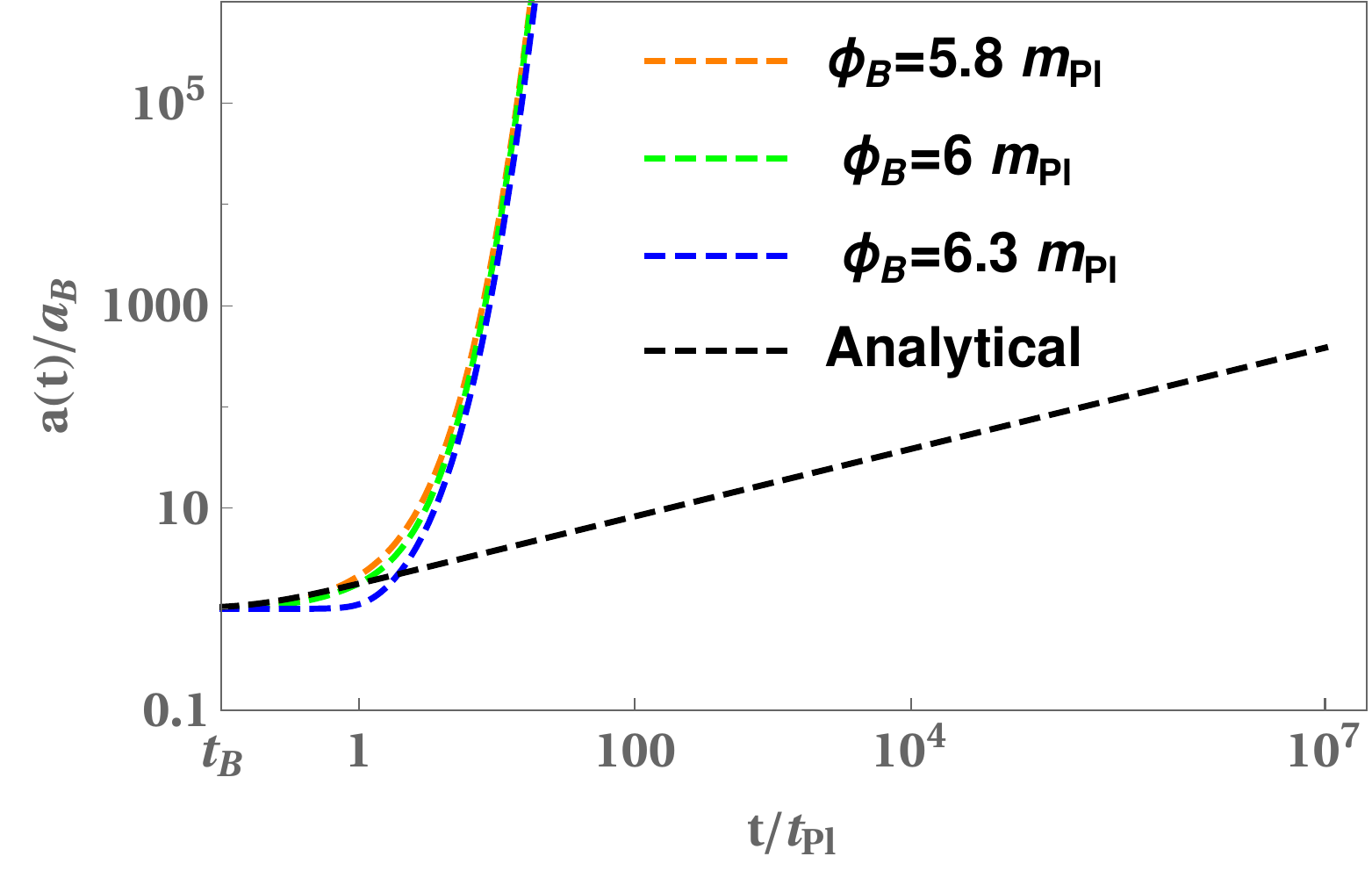}} & 
{\includegraphics[width=2.1in,height=1.6in,angle=0]{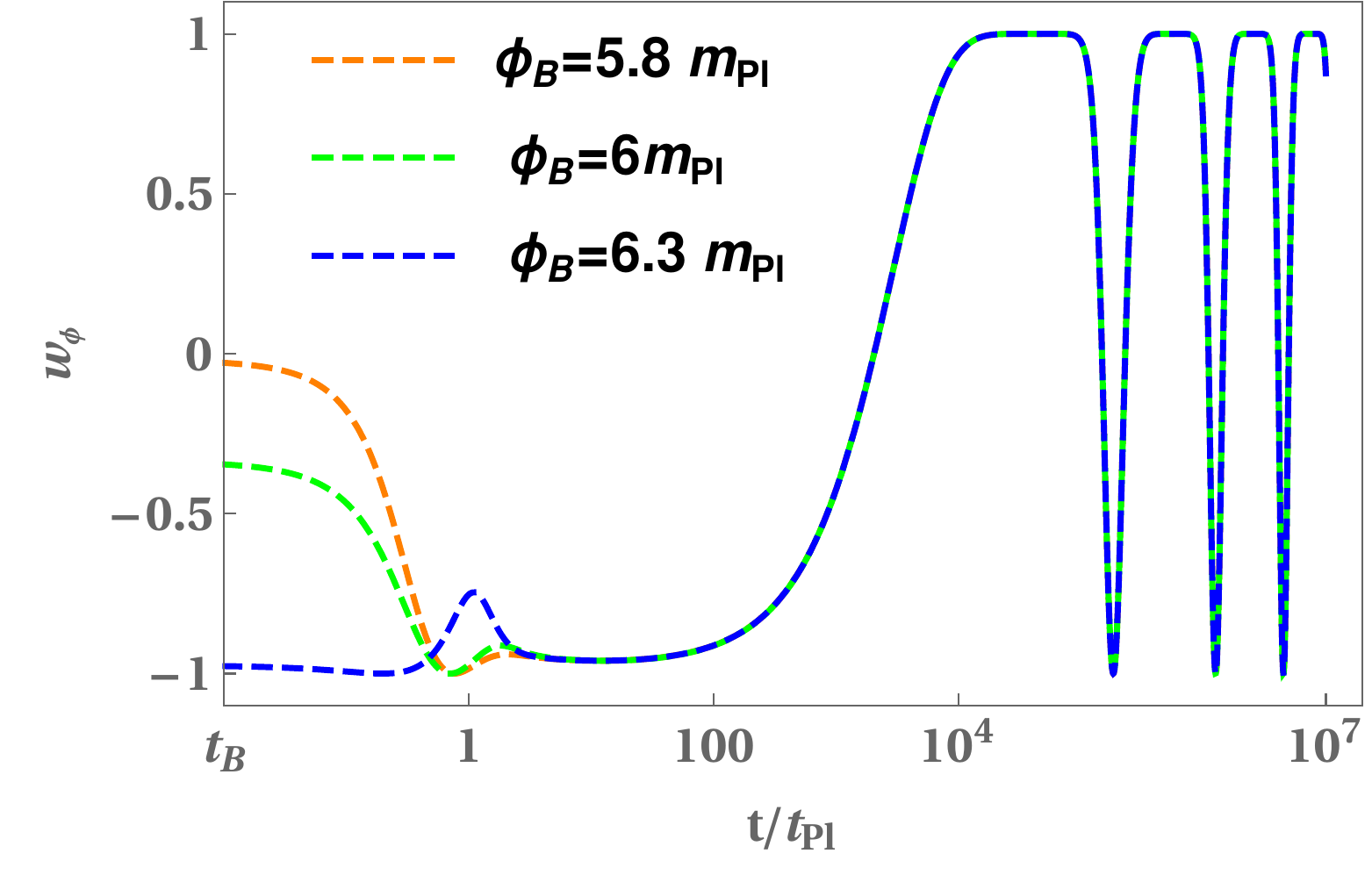}} & 
{\includegraphics[width=2.0in,height=1.6in,angle=0]{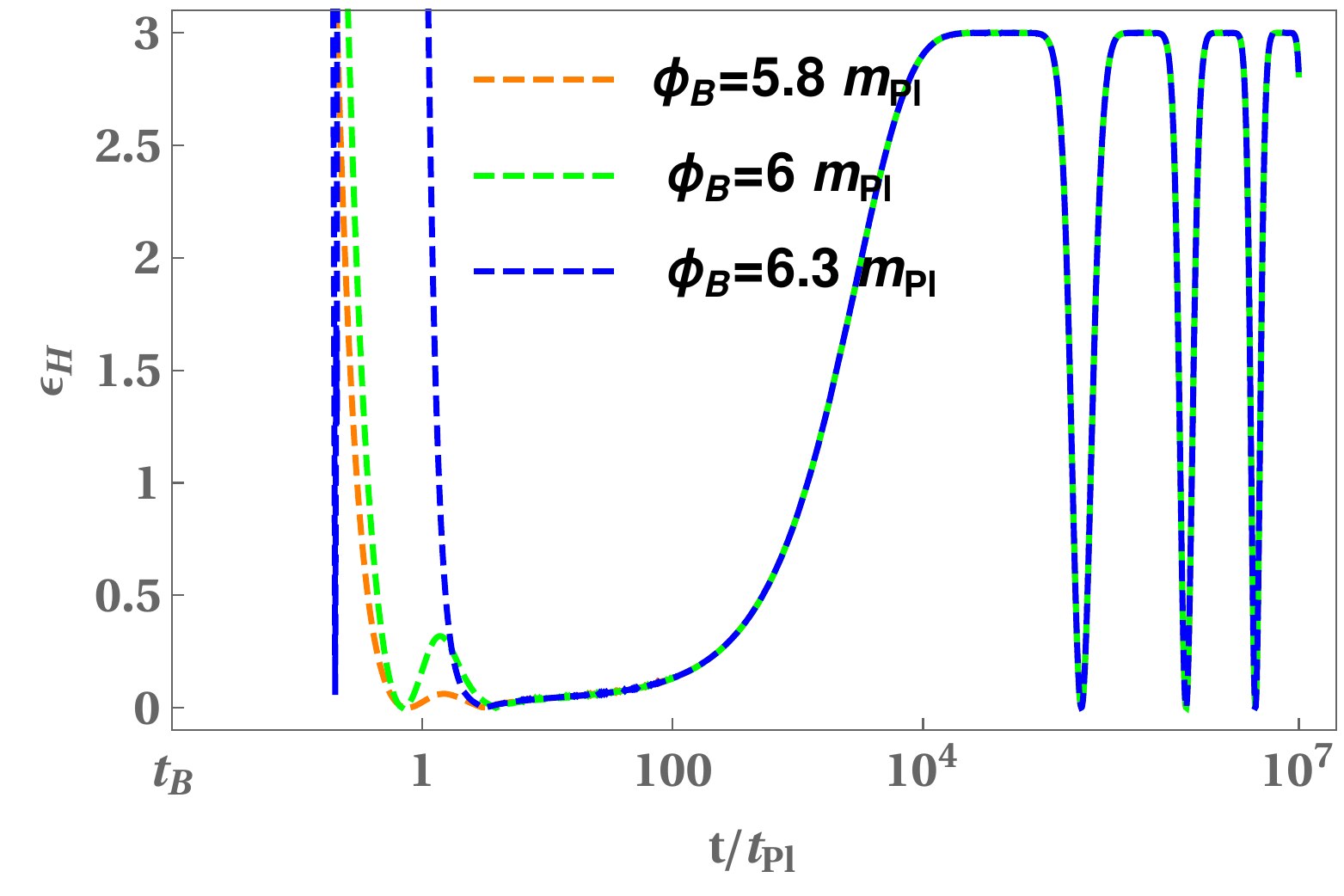}} 
\end{tabular}
\end{center}
\caption{The figure represents the results for Hilltop potential (\ref{eq:pot}) with  $p=4$ and $\dot{\phi}_B>0$. The numerical evolution of $a(t)$, $w(\phi)$ and $\epsilon_H$ is exhibited for the same set of KED (upper panels) and PED (lower panels) initial conditions of inflaton field at the bounce with $v=0.019 m_{Pl}$ and $V_0=2.632 \times 10^{-21} m_{Pl}^4$. The analytical solution of $a(t)$ (\ref{eq:a}) is also displayed in order to compare it with the numerical results.}
\label{fig:P4vpt1}
\end{figure*}

Let us first evolve the background equations (\ref{eq:Hub}) and (\ref{eq:ddphi}) with Hilltop potential (\ref{eq:pot}) numerically for $p=4$, $v=0.019 m_{Pl}$ and $V_0=2.632 \times 10^{-21} m_{Pl}^4$. 
In the top (bottom) panels of Fig. \ref{fig:P4vpt1}, we show the numerical results for a set of KED (upper panels) and PED (lower panels) initial conditions where the behavior of expansion factor $a(t)$, equation of state $w(\phi)$ and slow-roll parameter $\epsilon_H$ are depicted. From this figure, one can conclude that the scale factor is not consistent with the analytical solution (\ref{eq:a}) and does not provide the exponential expansion. Similarly, by looking the middle panels of Fig. \ref{fig:P4vpt1}, the equation of state does not give the slow-roll inflation. Therefore, in this case, neither KED nor PED initial conditions generate the slow-roll inflation.

\begin{table}
\caption{The table shows the range of $v$ with SR and NSR inflation for KED and PED initial conditions of inflaton field at the bounce.  }
\begin{center}
\begin{tabular}{l  l cc l}
\hline\\
$p$ & ~~ $v/M_{Pl}$ && Slow-roll inflation depends on the range of $\phi_B$  & 
\\  \cline{3-5} \\
& & \qquad KED (SR) & Existence of KED  & PED (SR)\\
& & \qquad (except subset) & subset (NSR) & \\
\hline\\
4 \qquad\qquad & $0 < v \leq 0.1$ & No & Yes & \qquad No \\
  \qquad\qquad  & $0.1 < v \leq 1 $  & Yes & Yes & \qquad Yes \\\\
5 \qquad\qquad & $0 < v \leq 1$ & No & No  & \qquad No\\
\\
\hline
\end{tabular}
\label{tab-v}
\end{center}
\end{table}

\begin{figure*}[tbp]
\begin{center}
\begin{tabular}{ccc}
{\includegraphics[width=2.1in,height=1.65in,angle=0]{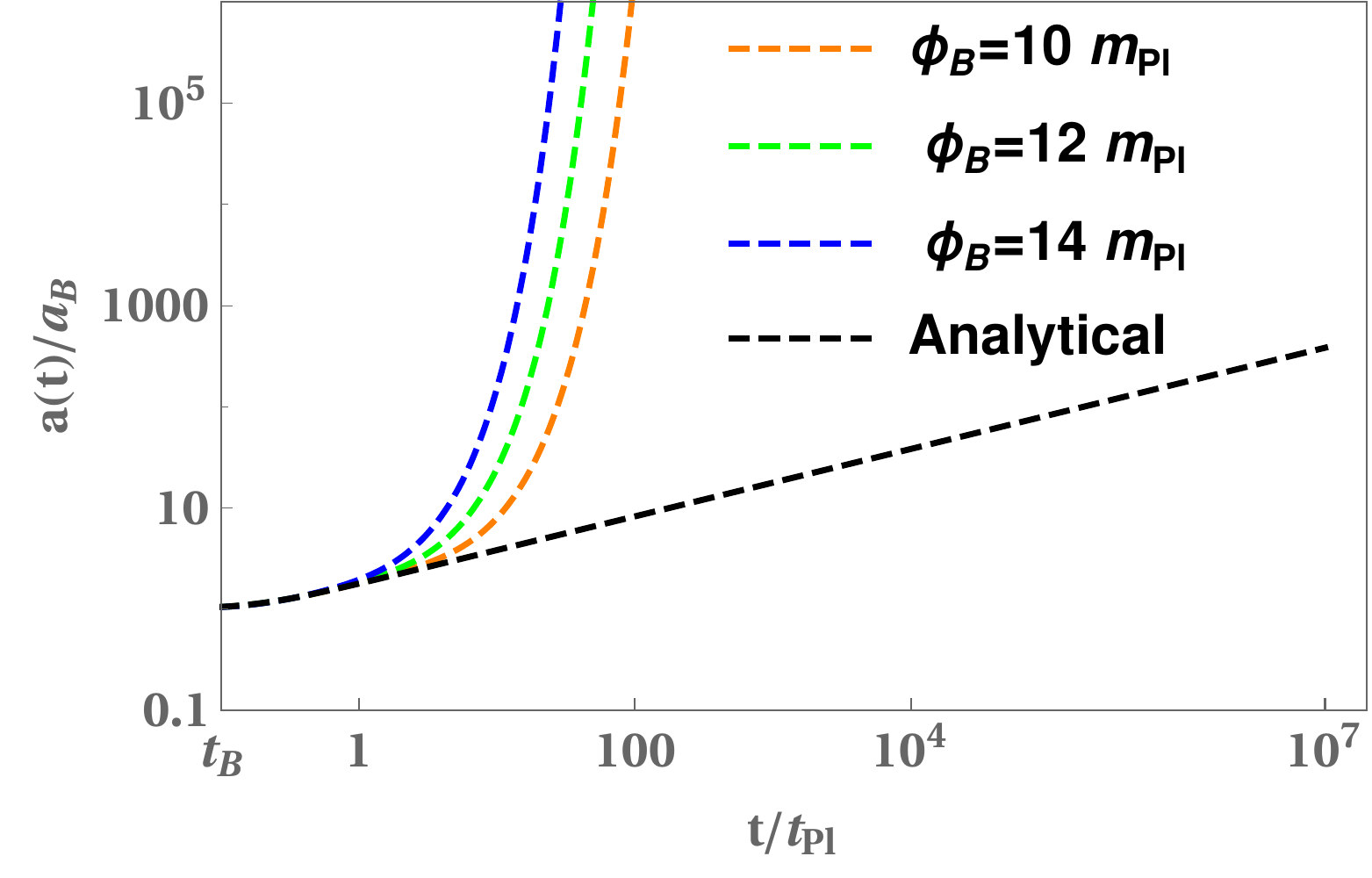}} &
{\includegraphics[width=2.1in,height=1.6in,angle=0]{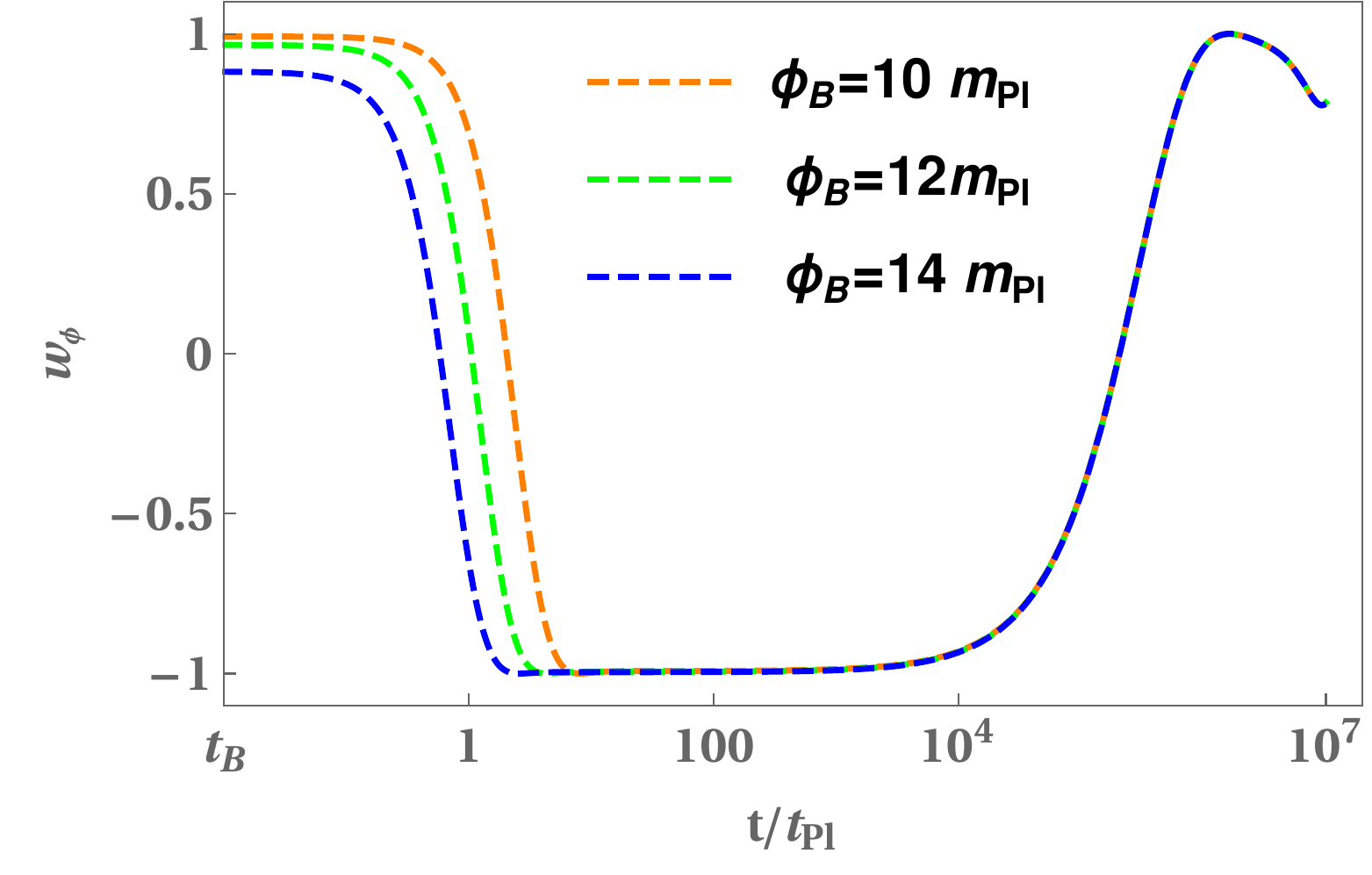}} &
{\includegraphics[width=2.0in,height=1.6in,angle=0]{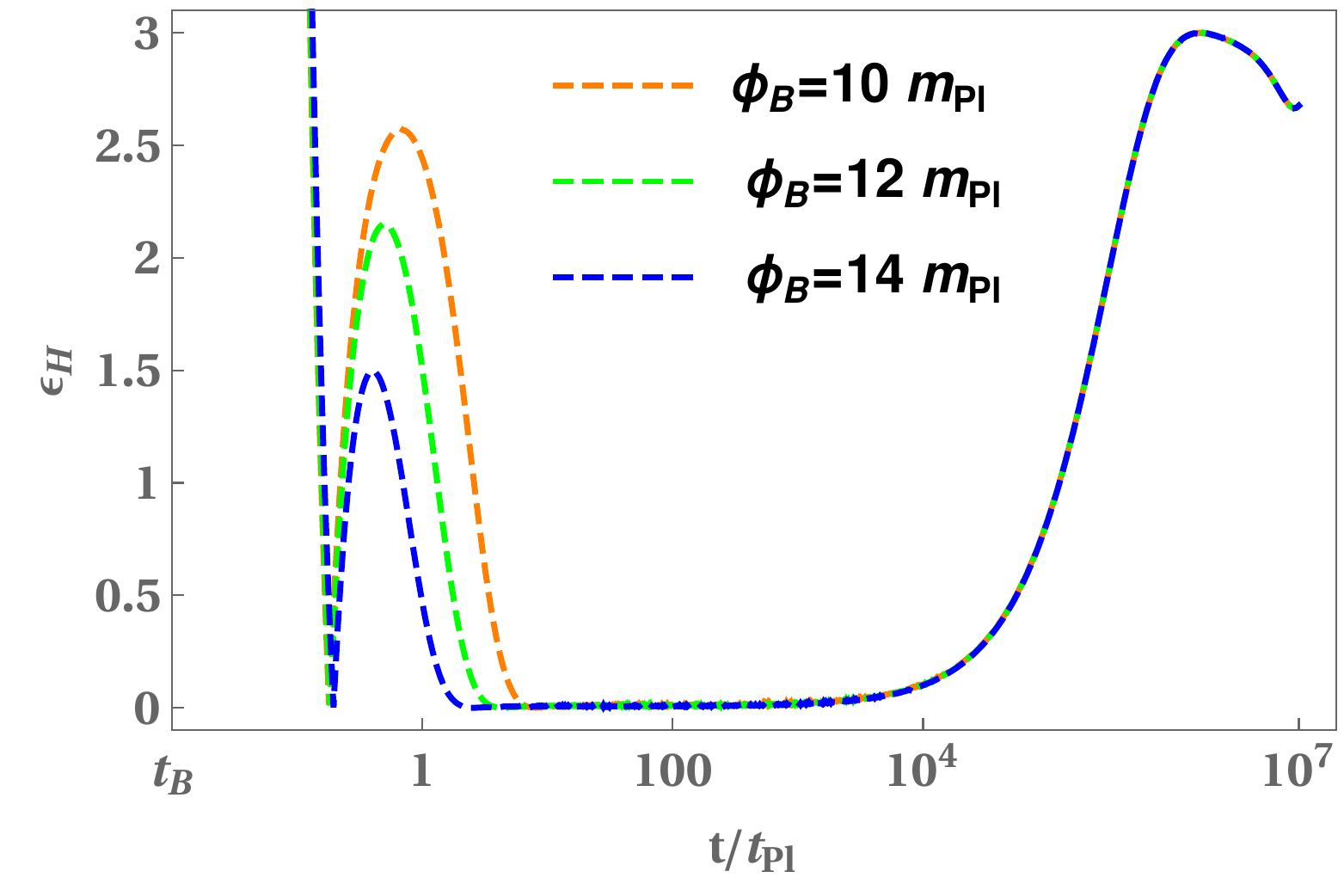}}
\\
{\includegraphics[width=2.1in,height=1.65in,angle=0]{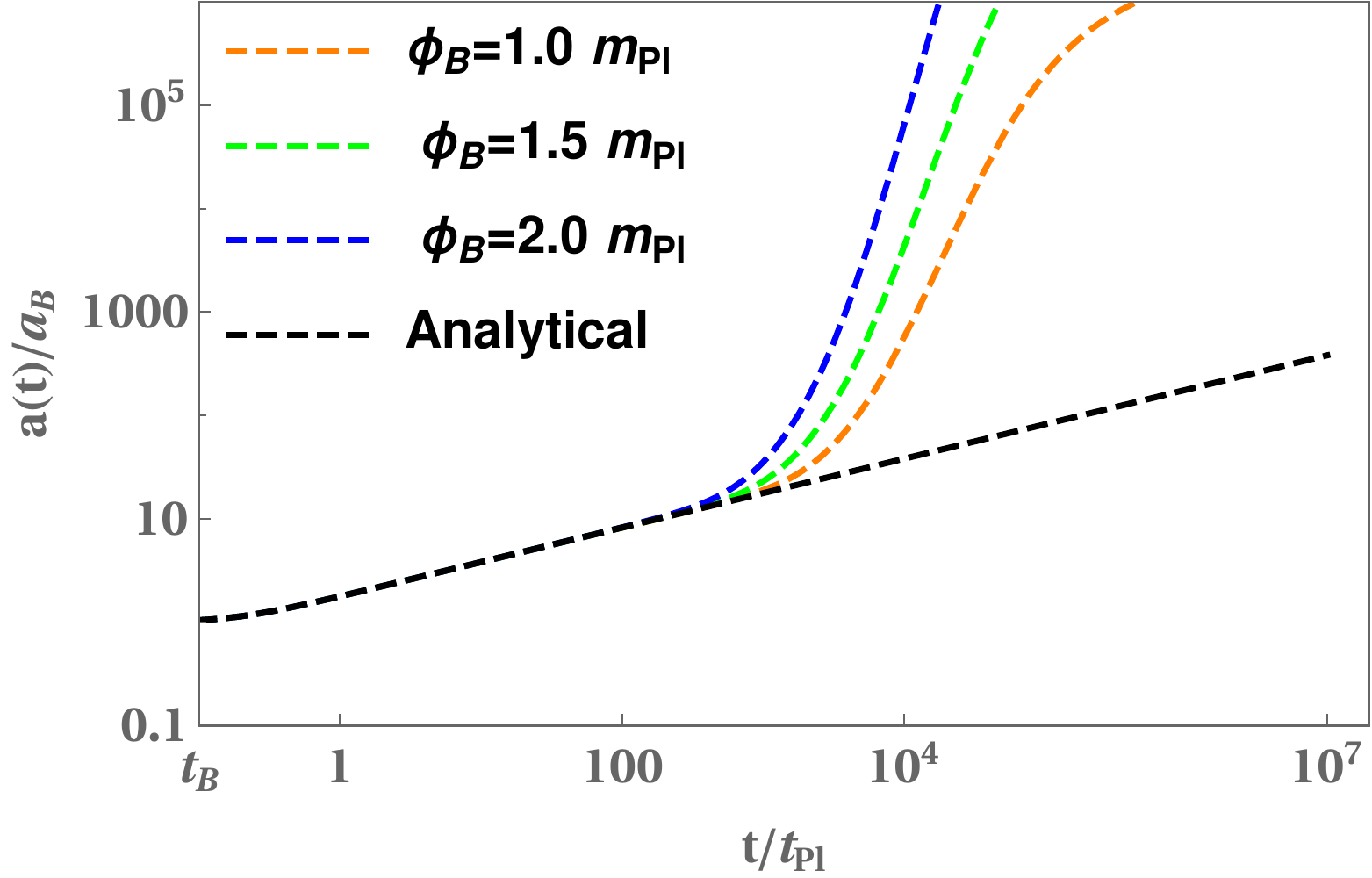}} &
{\includegraphics[width=2.1in,height=1.6in,angle=0]{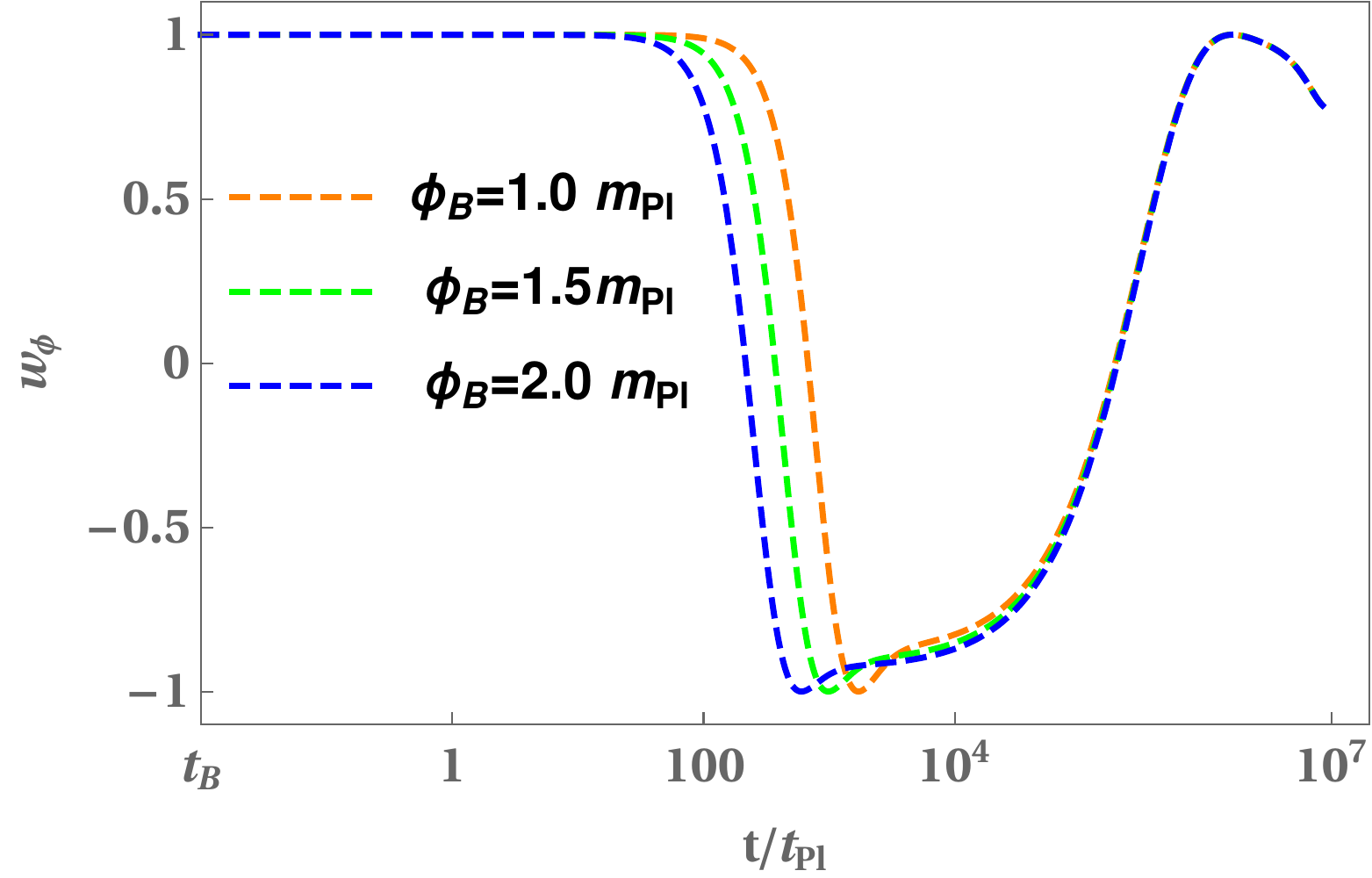}} &
{\includegraphics[width=2.0in,height=1.6in,angle=0]{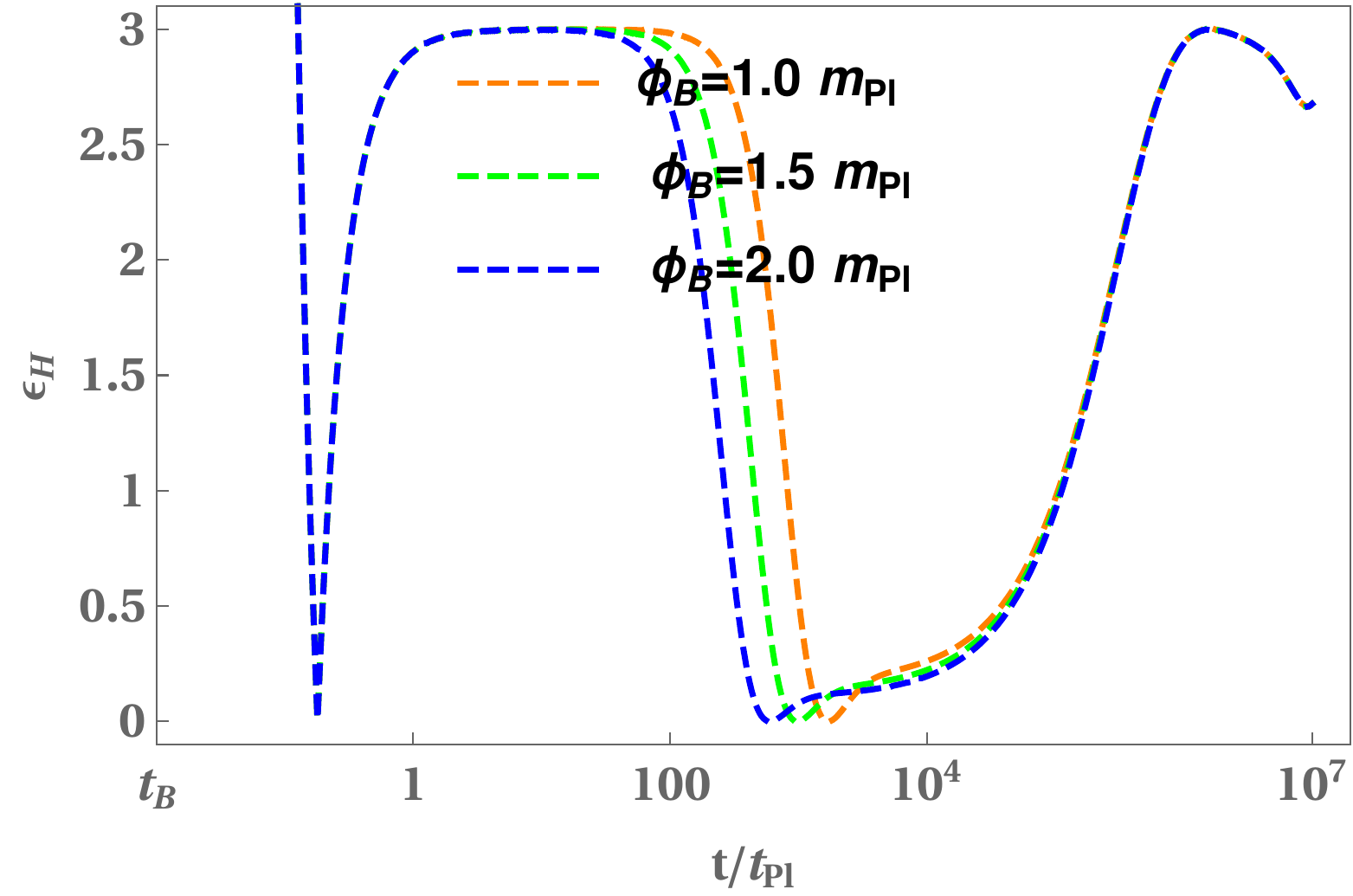}}
\\
{\includegraphics[width=2.1in,height=1.6in,angle=0]{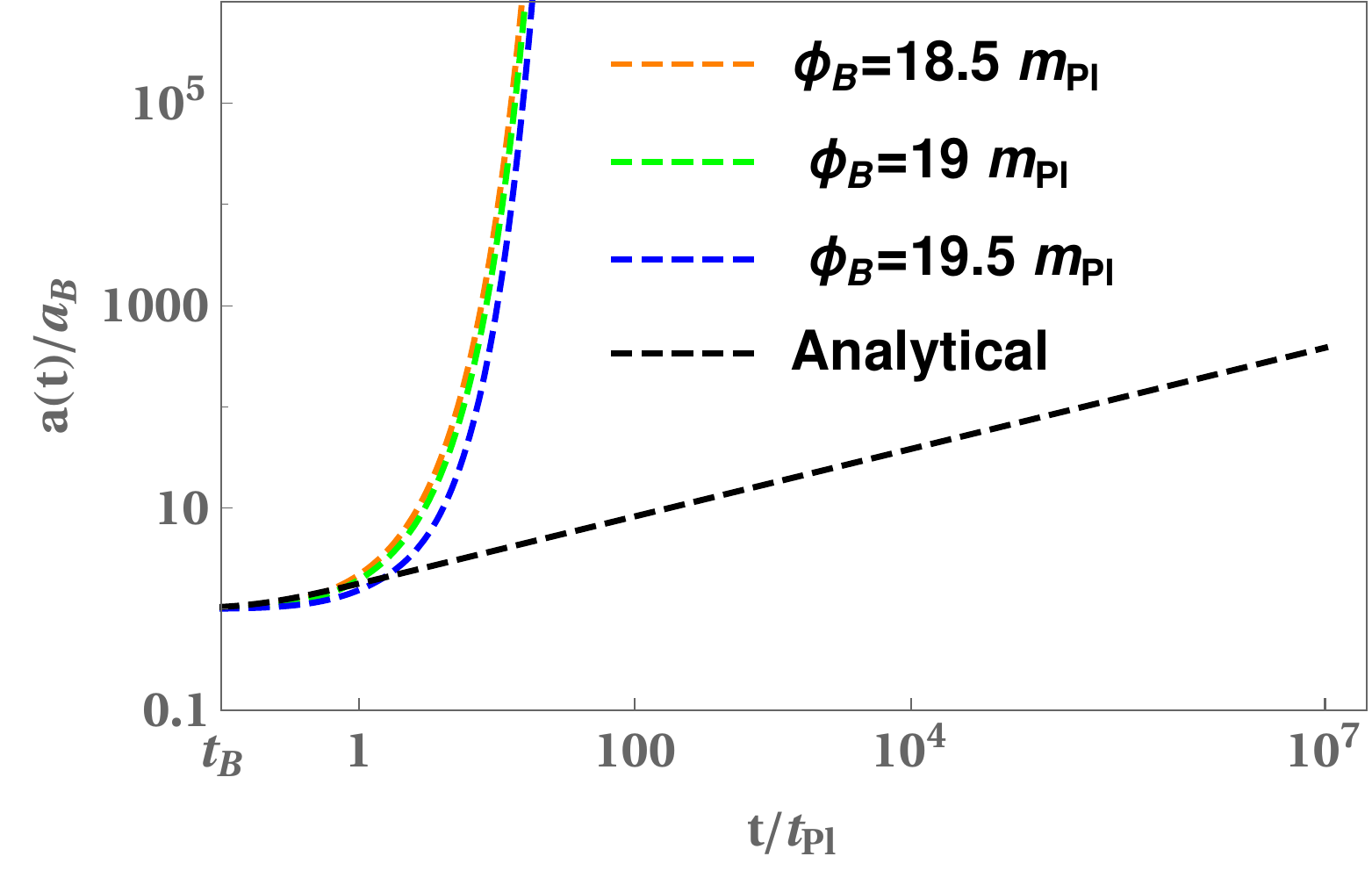}} & 
{\includegraphics[width=2.1in,height=1.6in,angle=0]{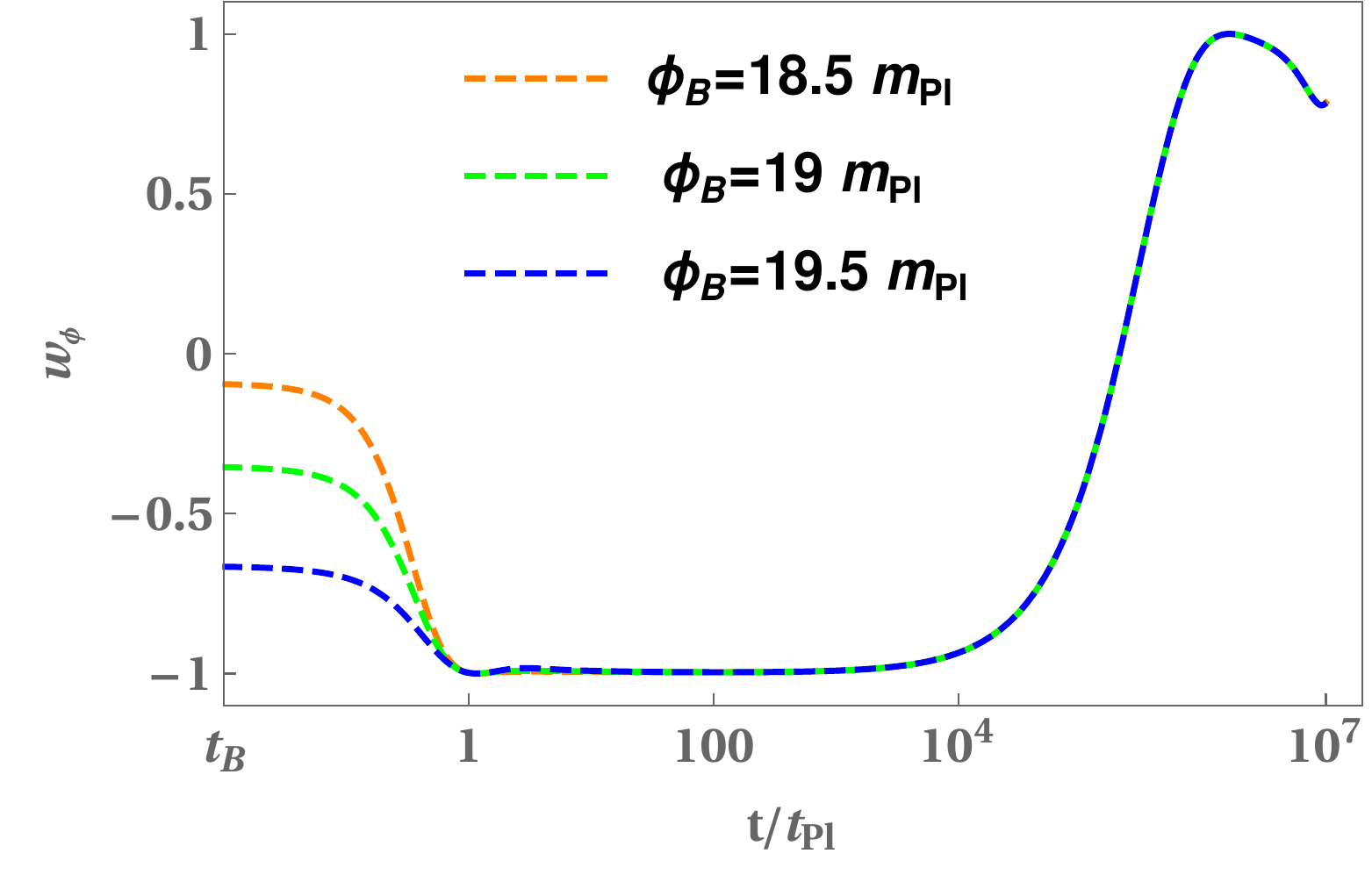}} & 
{\includegraphics[width=2.0in,height=1.6in,angle=0]{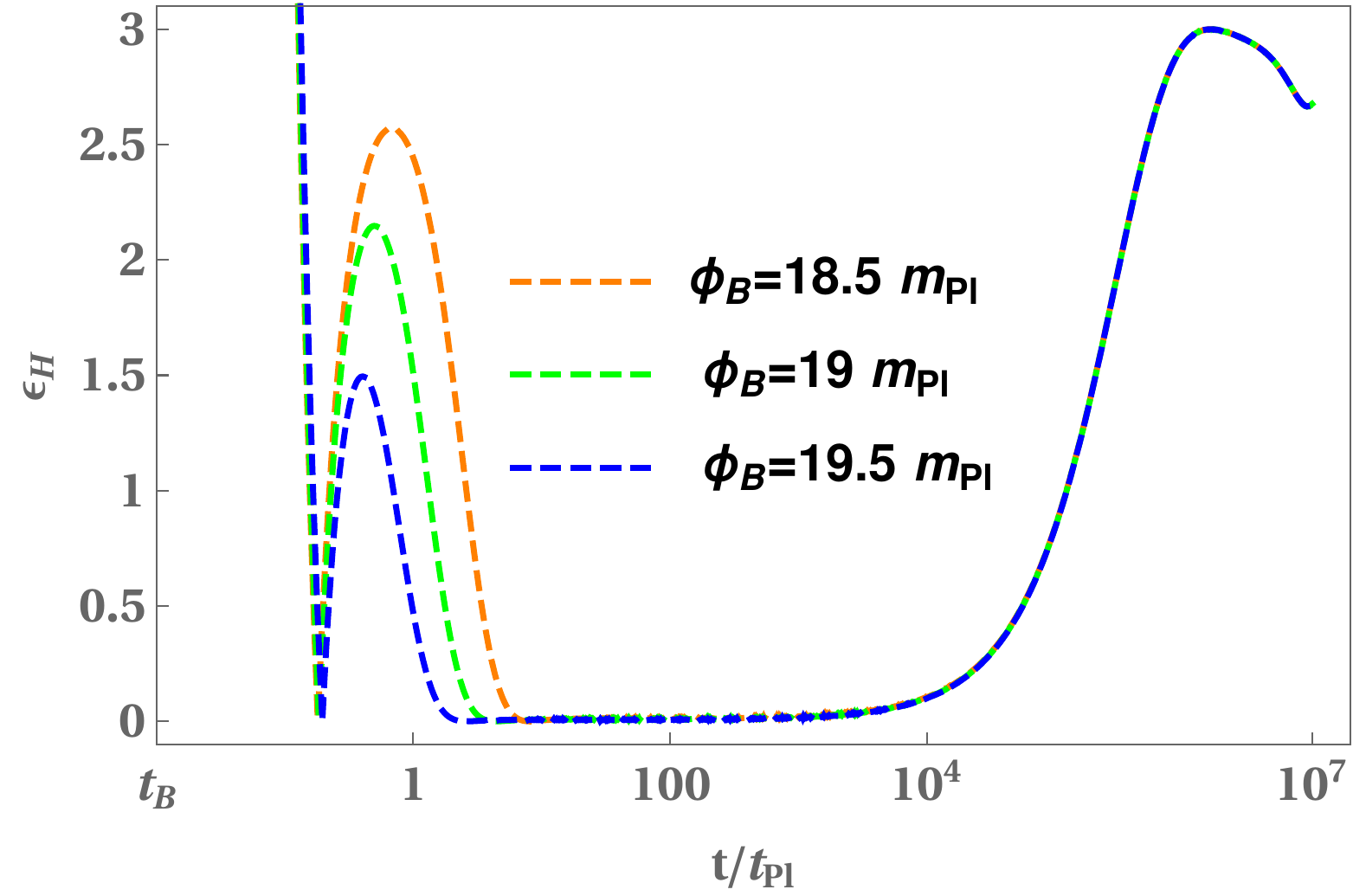}} 
\end{tabular}
\end{center}
\caption{ The figure shows the evolution of $a(t)$, $w(\phi)$ and $\epsilon_H$ for potential (\ref{eq:pot}) with $\dot{\phi}_B>0$. We choose  $p=4, v=0.19 m_{Pl}$ and $V_0=2.632 \times 10^{-17} m_{Pl}^4$ when plotting out the figure. Top and middle panels correspond to the KED initial conditions of inflaton field at the quantum bounce with SR and NSR, respectively whereas bottom panels are for PED initial conditions. In this figure, we get the desired slow-roll inflation for both KED (except small subset) and PED  initial conditions.}
\label{fig:P4v1}
\end{figure*}

\begin{table}
\caption{Table for SR and NSR inflation for different values of $v$ with $\dot{\phi_B}>0$. The value of $V_0$ for each $p$ and $v$ is given by equation (\ref{eq:pvV0}). The symbol $\forall$ represents for all. }
\begin{center}
\begin{tabular}{l  l cc l}
\hline\\
$p$ &  $v/M_{Pl}$ & \qquad KED (SR) & \qquad~~ Subset of KED  & \qquad\qquad PED (SR)\\
& & \qquad (except subset) &  \qquad (NSR) & \\
\hline\\
4 \qquad\qquad & ~0.1 & No & \qquad $\forall ~ \phi_B$ & \qquad\qquad~~ No \\
  \qquad\qquad  & ~1 & $5 \leq \phi_B \leq 18.3$ & \qquad $0 \leq \phi_B < 5$ & \qquad $18.3 < \phi_B \leq 19.9$ \\\\
5 \qquad\qquad & ~0.1 & No & \qquad $\forall ~ \phi_B$ & \qquad\qquad~~ No \\
  \qquad\qquad  & ~1 & No & \qquad $\forall ~ \phi_B$ & \qquad\qquad~~ No \\\\
\hline
\end{tabular}
\label{tab-phiB}
\end{center}
\end{table}

\begin{center}
\begin{table}
\caption{This table represents the Hilltop potential (\ref{eq:pot}) with $p=4$,  $v=0.19m_{Pl}$ and $V_0=2.632 \times 10^{-17} m_{pl}^4$. We show the number of $e$-foldings $N_{inf}$ and other important parameters of inflation. }
\resizebox{0.65\textwidth}{!}{%
\begin{tabular}{l  l  c  c  c  l}
\hline
$\phi_B/m_{Pl}$ \qquad  & \qquad Inflation \qquad & \qquad $t/t_{pl}$ \qquad & \qquad $\epsilon_H$ \qquad & \qquad $w(\phi)$ \qquad & \qquad $N_{inf}$ \\
\hline\\
5 & \qquad start & \qquad 26.048 & \qquad 0.999 & \qquad $-1/3$ & \qquad 56.11\\
& \qquad slow-roll & \qquad 69.294 & \qquad $2.400 \times 10^{-5}$ & \qquad $-1.0$ &\\
& \qquad end & \qquad $1.198\times 10^5$ & \qquad 1.000 & \qquad $-1/3$ &\\\\
5.22 & \qquad start & \qquad 22.847 & \qquad 0.999 & \qquad $-1/3$ & \qquad 60.01\\
& \qquad slow-roll & \qquad 61.318 & \qquad $1.612 \times 10^{-5}$ & \qquad $-1.0$ &\\
& \qquad end & \qquad $1.201\times 10^5$ & \qquad 0.999 & \qquad $-1/3$ &\\\\
7 & \qquad start & \qquad 8.928 & \qquad 1.000 & \qquad $-1/3$ & \qquad 95.92\\
& \qquad slow-roll & \qquad 25.557 & \qquad $2.171 \times 10^{-5}$ & \qquad $-1.0$ &\\
& \qquad end & \qquad $1.213\times 10^5$ & \qquad 0.999 & \qquad $-1/3$ &\\\\
10 & \qquad start & \qquad 2.608 & 0.991 & \qquad $-1/3$ & \qquad 177.08\\
& \qquad slow-roll & \qquad 8.106 & \qquad $1.145 \times 10^{-4}$ & \qquad $-1.0$ &\\
& \qquad end & \qquad $1.223\times 10^5$ & \qquad 1.000 & \qquad $-1/3$ &\\\\
\hline
\end{tabular}
}
\label{tab}
\end{table}
\end{center}


Next, we choose  $p=4$, $v=0.19 m_{Pl}$ and $V_0=2.632 \times 10^{-17} m_{Pl}^4$. Further, we numerically evolve background  equations (\ref{eq:Hub}) and (\ref{eq:ddphi}) with Hilltop potential (\ref{eq:pot}). The results of numerical evolution for KED (top and middle panels) and PED (bottom Panels) initial conditions are displayed in Fig. \ref{fig:P4v1}. First, let us discuss the set of KED initial values (only for top panels) in the bouncing phase, the behavior of numerical evolution of $a(t)$ is universal because it neither depends on the initial values of inflaton field nor on the potential, and is in good agreement with the analytical solution (\ref{eq:a}). This happens mainly due to the fact that the contribution of potential is very small as compared to the kinetic one in the entire bouncing regime. As a result, it shows negligible effects on the evolution of background. By looking on the evolution of equation of state $w(\phi)$ (top panel), one can infer that the evolution of background is categorized into three different phases such as bouncing, transition and slow-roll inflation. On comparison with the period of three regimes, the transition phase has a very small period in contrast with the bouncing and slow-roll phases. During the bouncing regime, $w(\phi) \simeq +1$, in the transition regime, it reduces from $+1~  (t/t_{Pl} \approx t_B)$ to $-1 ~ (t/t_{Pl} \approx 10)$, while in the slow-roll regime, it is almost $-1$ until the end of the slow-roll inflation. Similarly, one can infer from the upper right panel, the slow-roll parameter $\epsilon_H>1$ in the bouncing phase, decreases from $\epsilon_H>1$ to $\epsilon_H \approx 0$ during the transition regime, and remains $\epsilon_H \approx 0$ untill the end of slow-roll inflation. A subset of KED initial conditions of inflaton field is also exist that does not provide the SR inflation which is depicted in the middle panels of Fig. (\ref{fig:P4v1}), and the range of this subset in terms of $\phi_B$ is shown in Table \ref{tab-phiB}.

Second, we discuss the case of PED initial conditions of inflaton field at the quantum bounce (bottom panels), the numerical evolution of scale factor $a(t)$ shows that the universality is lost, bouncing and transition regimes do not exist any more. However, the slow-roll inflation can still be achieved, see the bottom panels of Fig. \ref{fig:P4v1}. In Table \ref{tab}, we display the various inflationary parameters, namely,  $\epsilon_H$, $w(\phi)$ and $N_{inf}$ etc. The desired slow-roll inflation is produced for different values of $\phi_B$, and the number of $e$-folds are obtained. For the successful inflation, at least 60 $e$-folds are required that are presented in Table \ref{tab}. From the table, one can notice that $N_{inf}$ increases as the initial values of inflaton field grows at the bounce.

\begin{figure*}[tbp]
\begin{center}
\begin{tabular}{ccc}
{\includegraphics[width=2.1in,height=1.65in,angle=0]{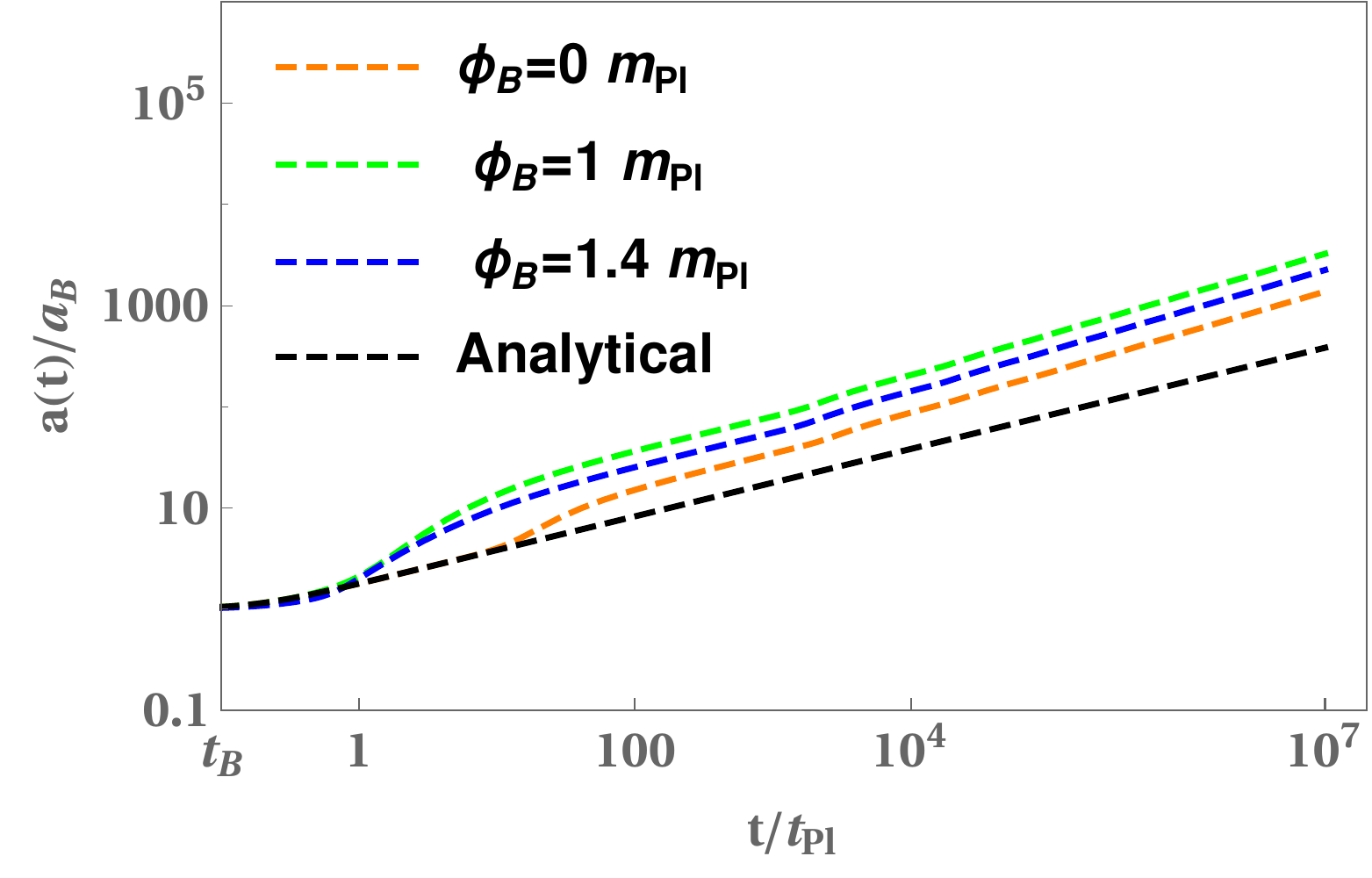}} &
{\includegraphics[width=2.1in,height=1.6in,angle=0]{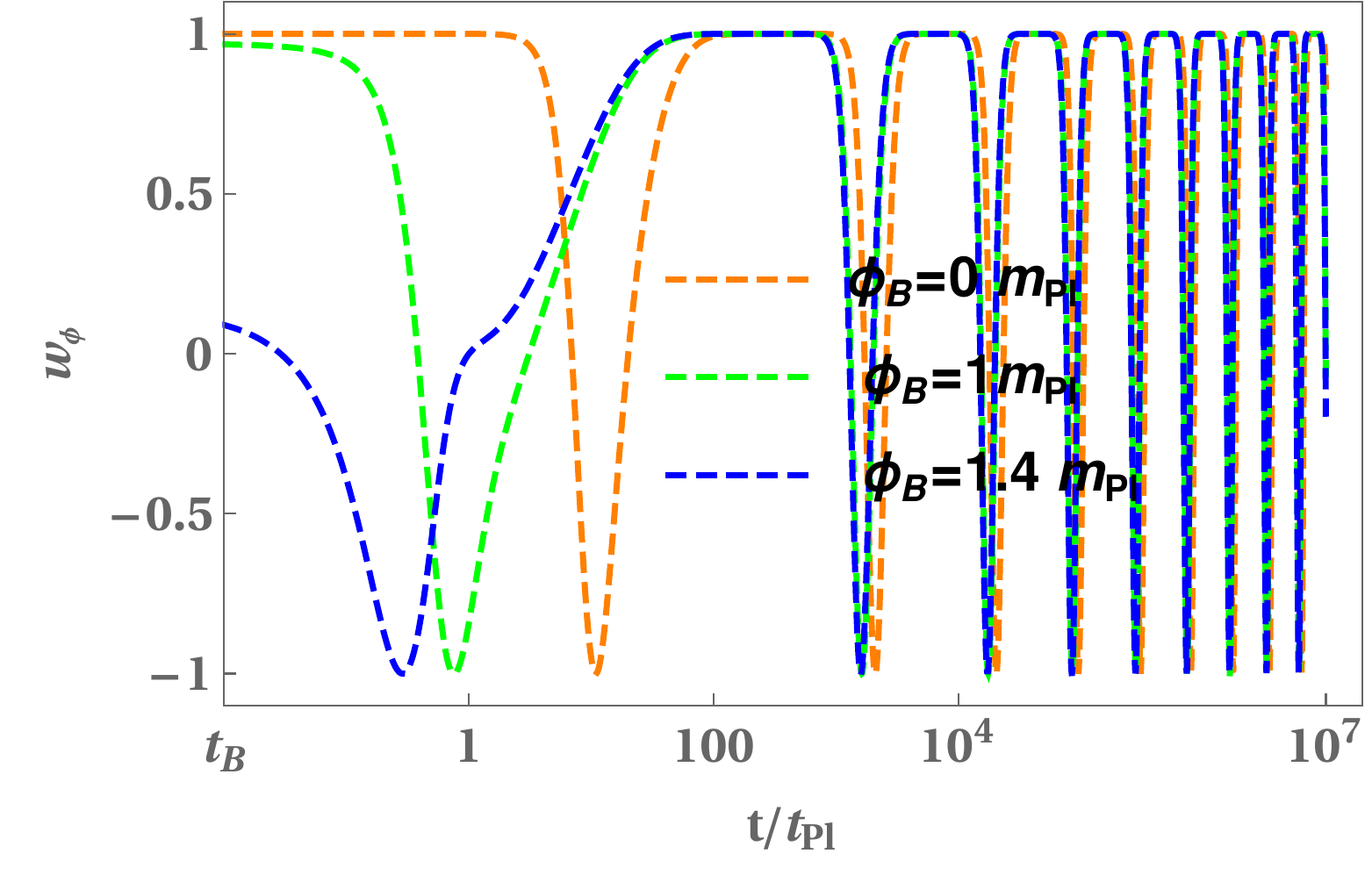}} &
{\includegraphics[width=2.0in,height=1.6in,angle=0]{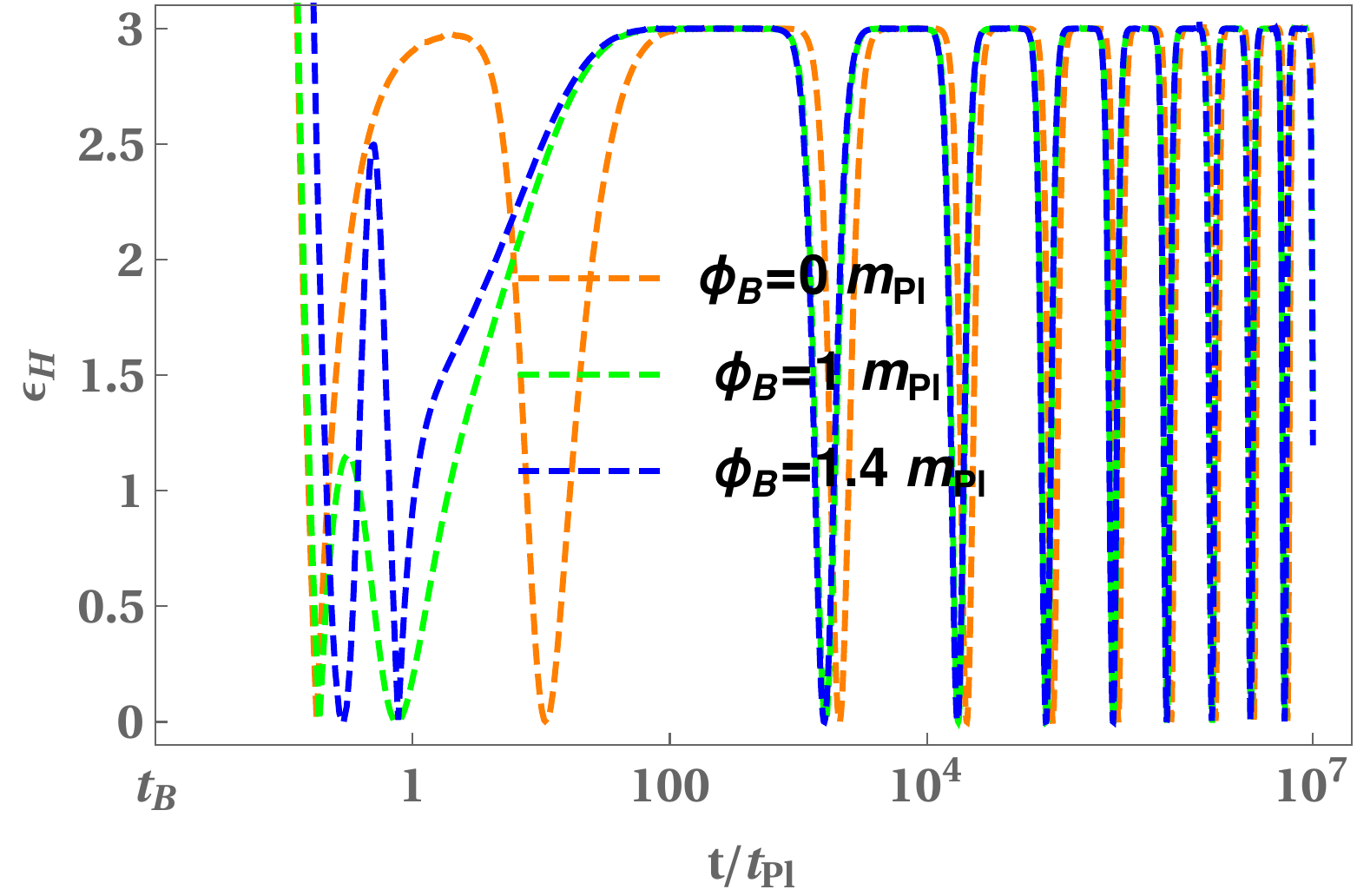}}
\\
{\includegraphics[width=2.1in,height=1.6in,angle=0]{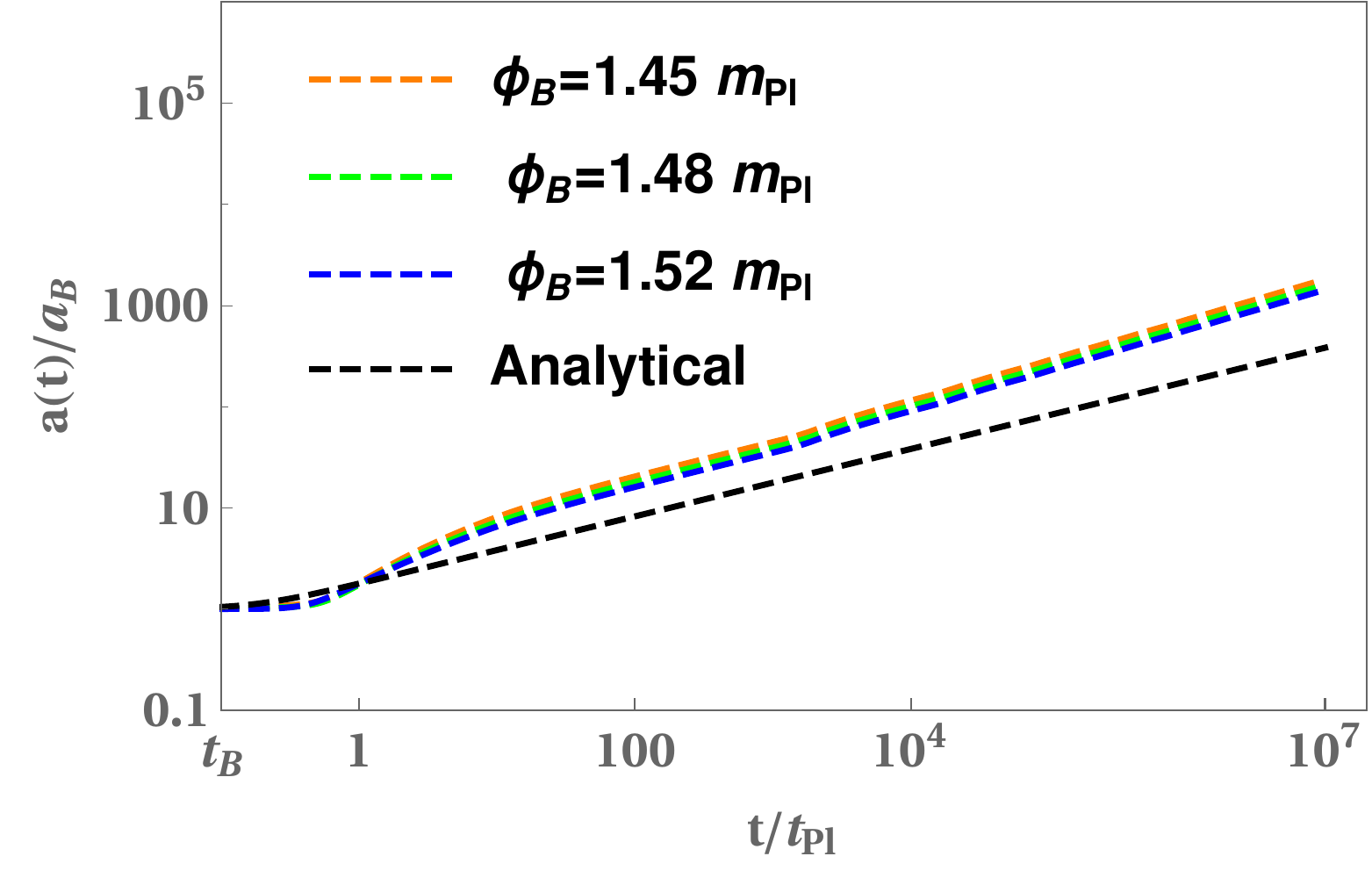}} & 
{\includegraphics[width=2.1in,height=1.6in,angle=0]{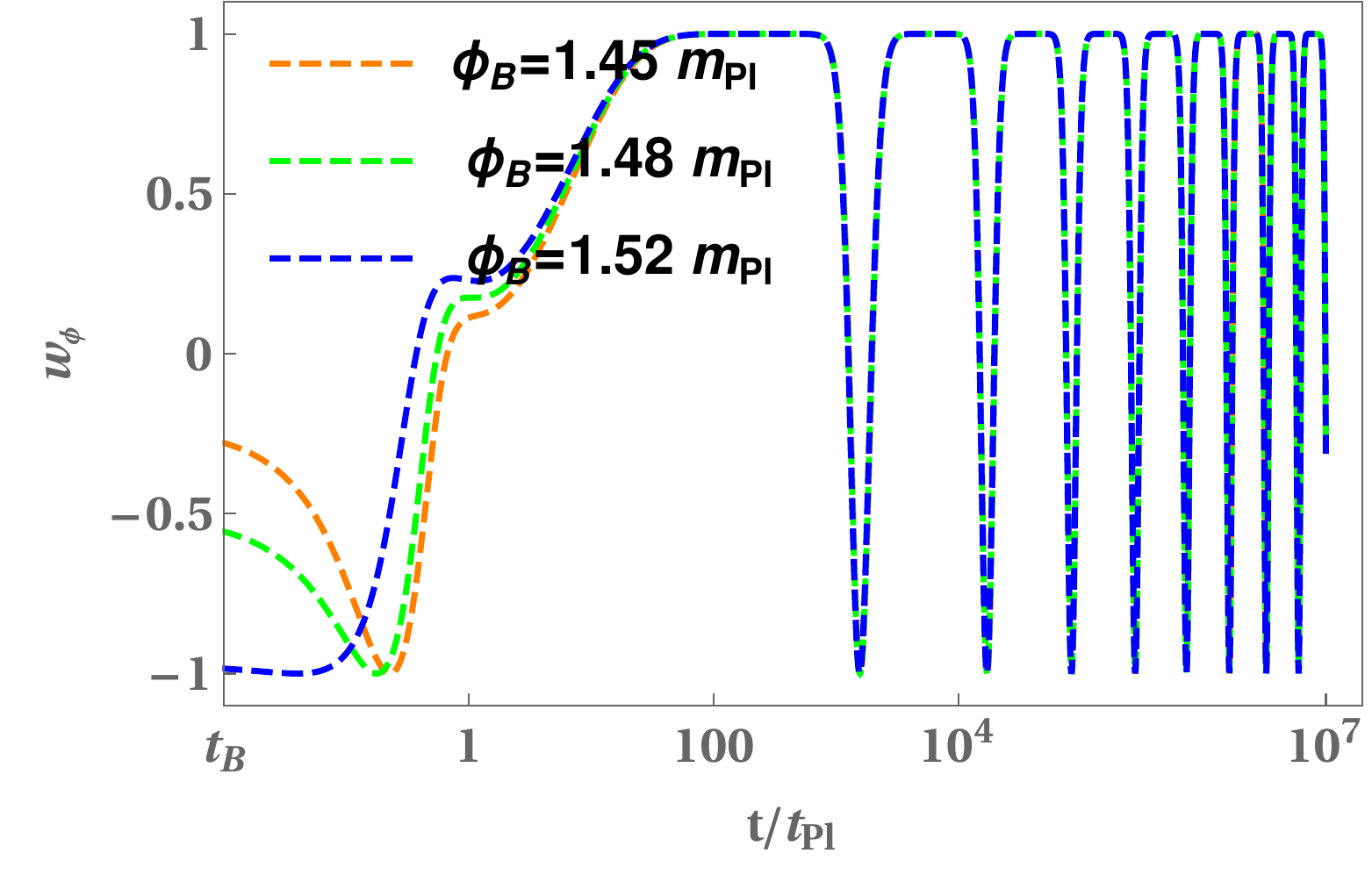}} & 
{\includegraphics[width=2.0in,height=1.6in,angle=0]{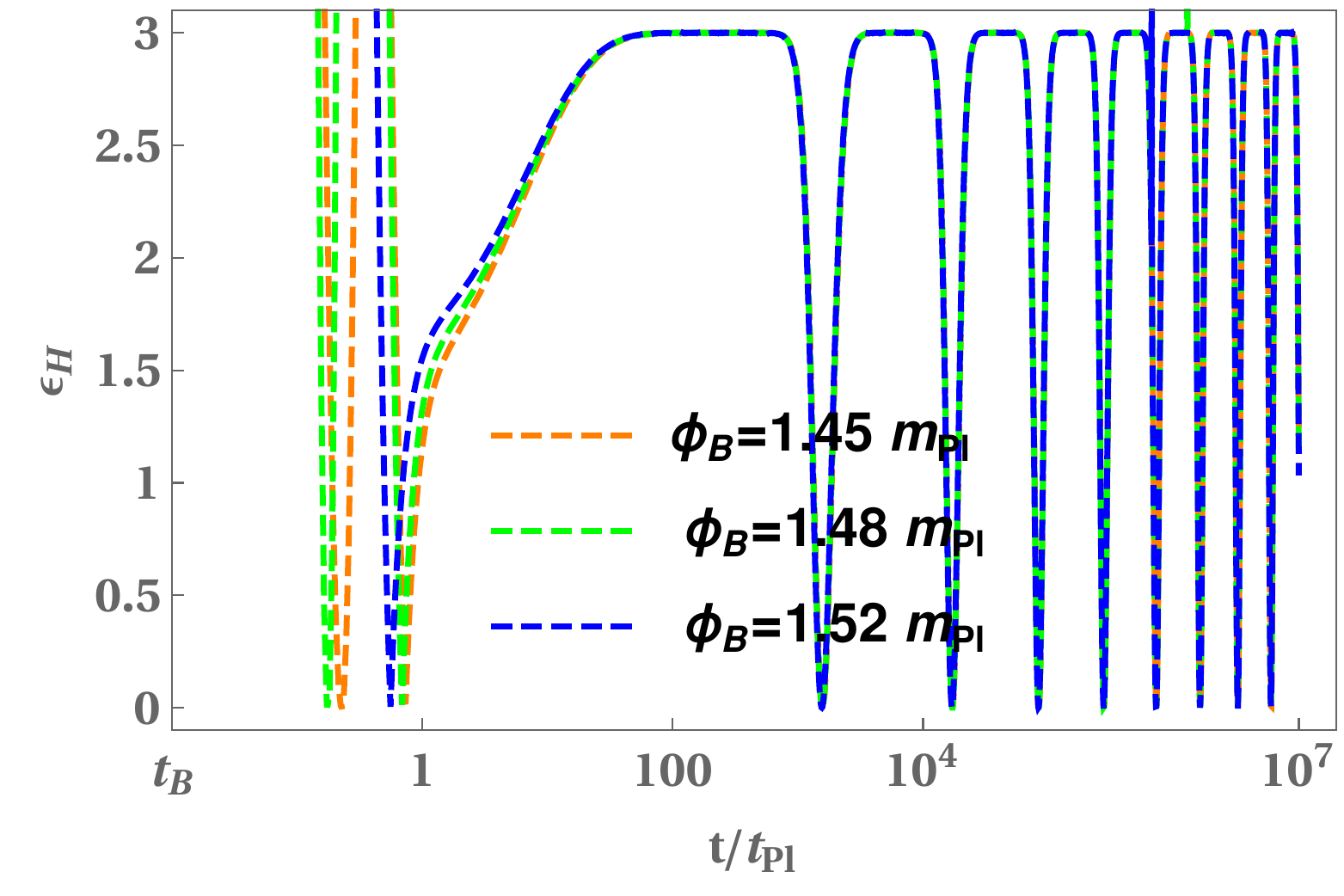}} 
\end{tabular}
\end{center}
\caption{ This figure demonstrates the numerical results for Hilltop potential with $p=5, v=0.019 m_{Pl}$ and $V_0=3.577 \times 10^{-20} m_{Pl}^4$. The numerical evolution of $a(t)$ does not show exponential expansion, $w(\phi)$ and $\epsilon_H$ are not close to $-1$ and less than unity, respectively during the whole evolution. Thereofre, by looking at the evolution of $a(t)$, $w(\phi)$ and $\epsilon_H$, we conclude that the slow-roll inflation is not achieved.}
\label{fig:P5vpt1}
\end{figure*}

\begin{figure*}[tbp]
\begin{center}
\begin{tabular}{ccc}
{\includegraphics[width=2.1in,height=1.65in,angle=0]{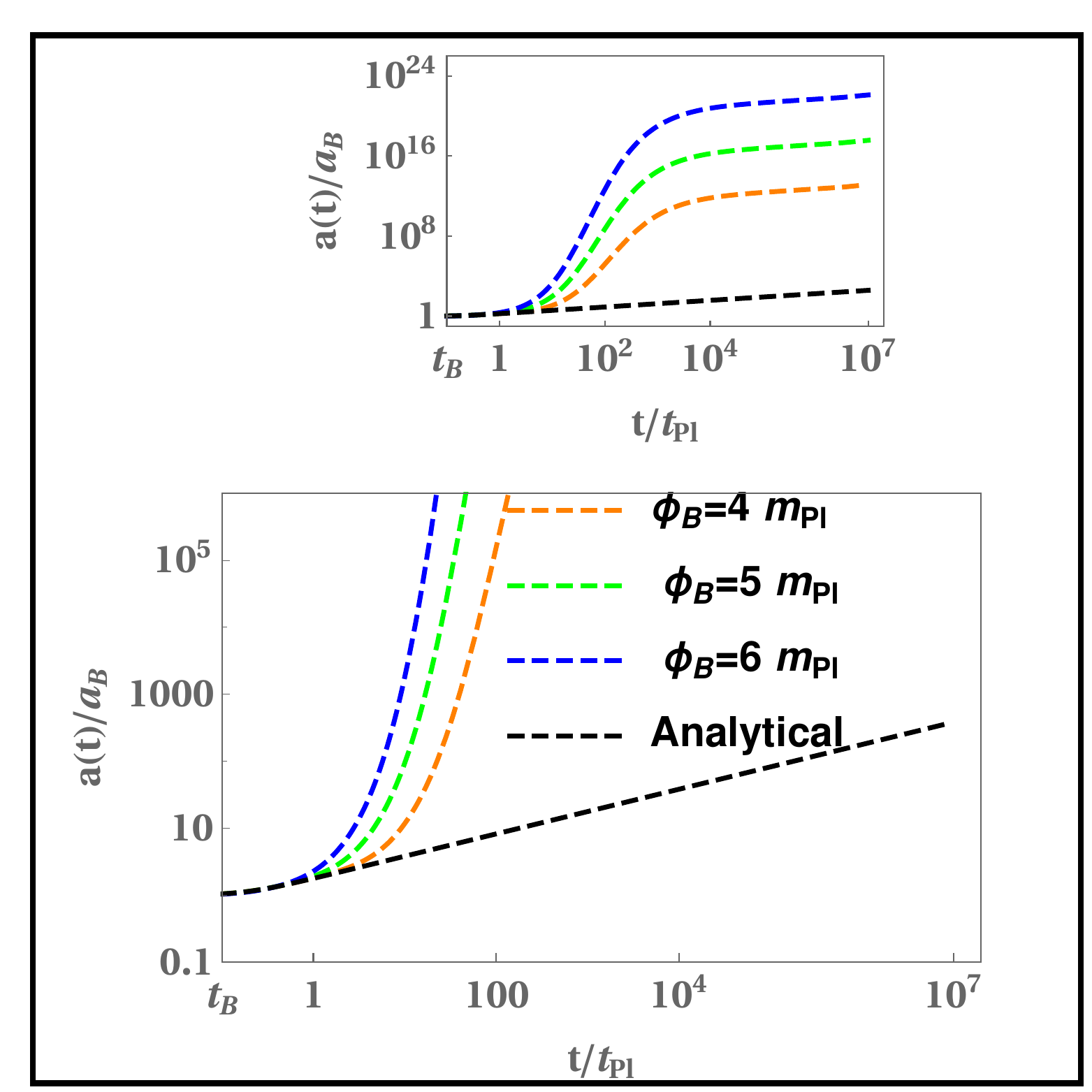}} &
{\includegraphics[width=2.1in,height=1.6in,angle=0]{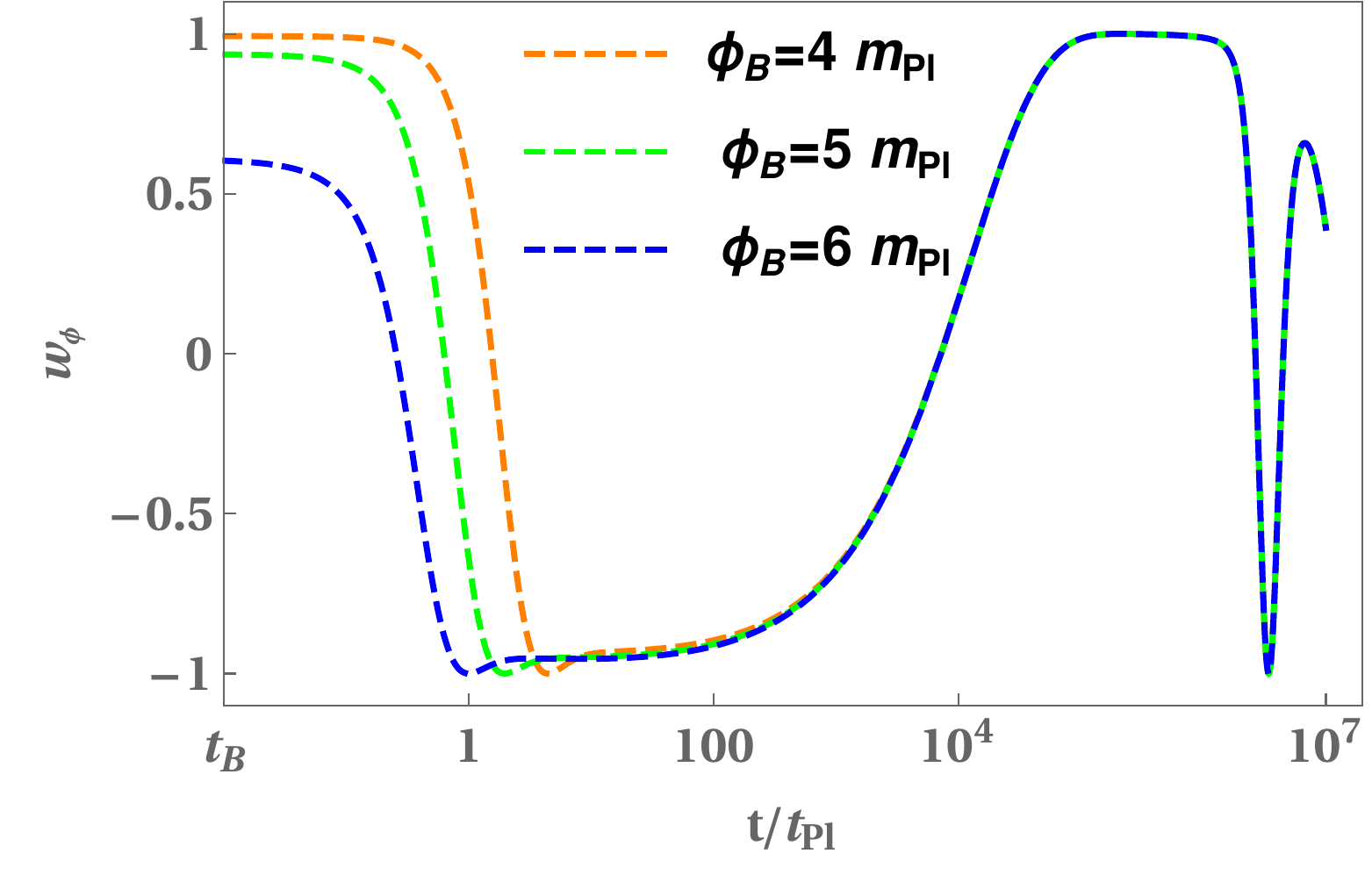}} &
{\includegraphics[width=2.0in,height=1.6in,angle=0]{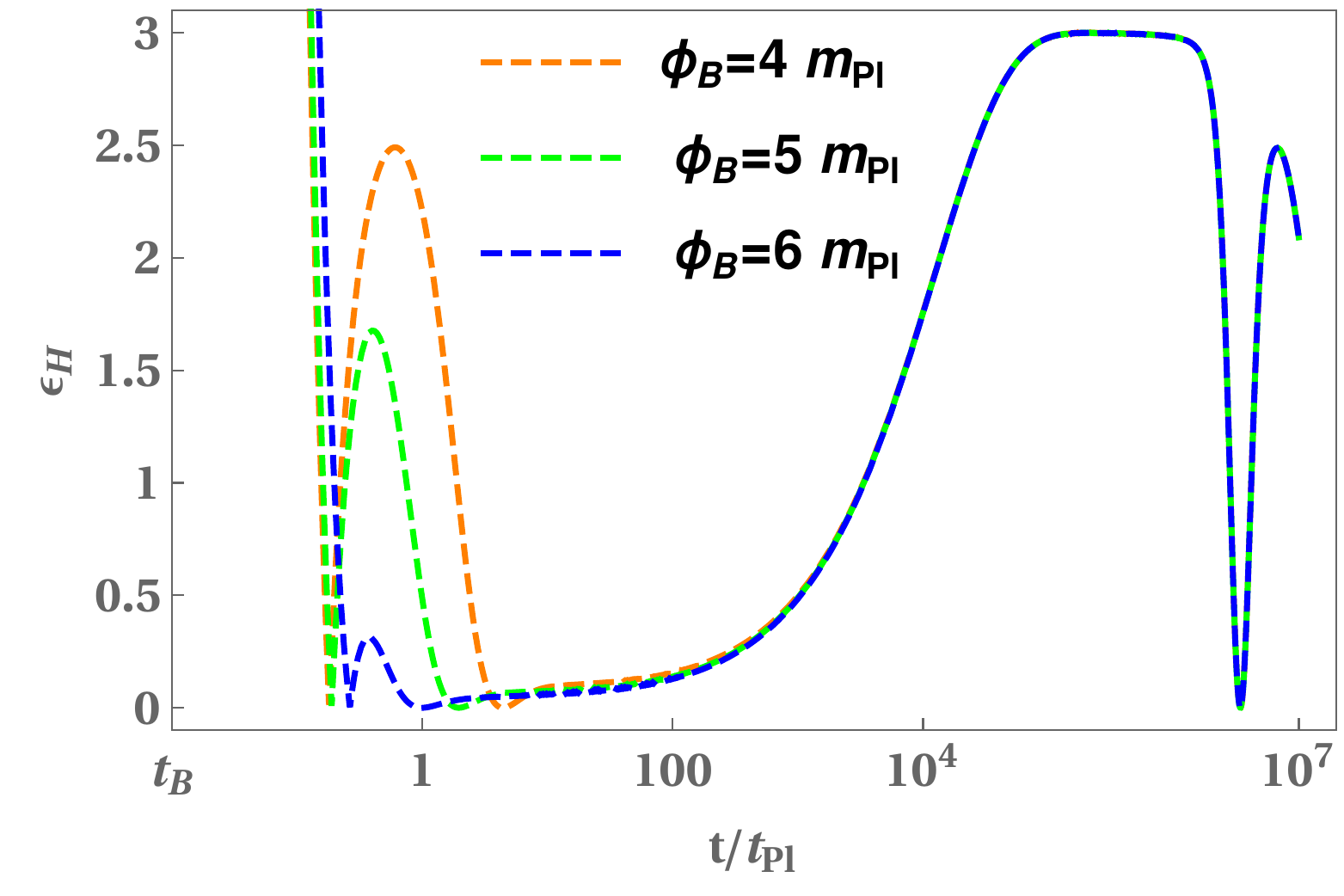}}
\\
{\includegraphics[width=2.1in,height=1.65in,angle=0]{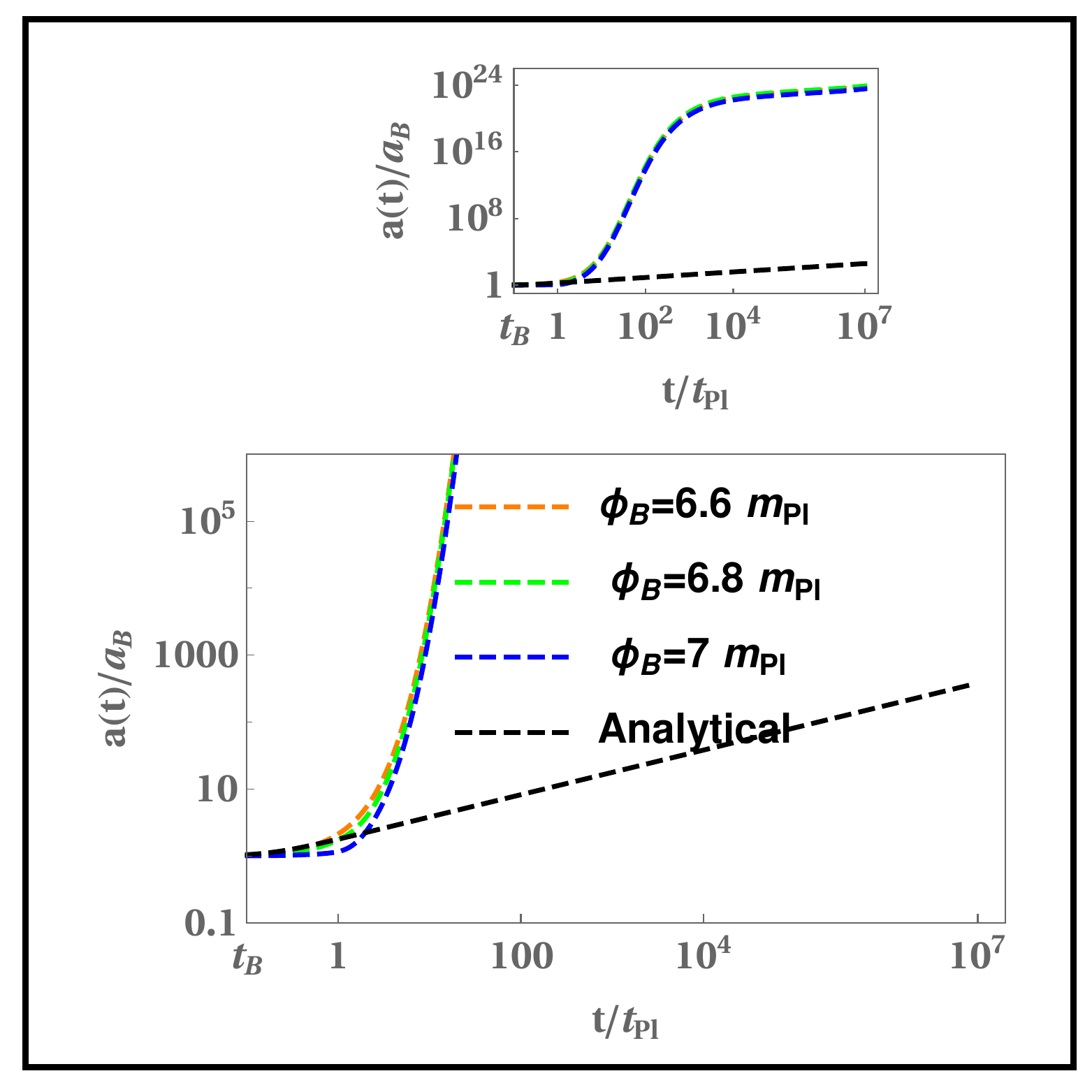}} & 
{\includegraphics[width=2.1in,height=1.6in,angle=0]{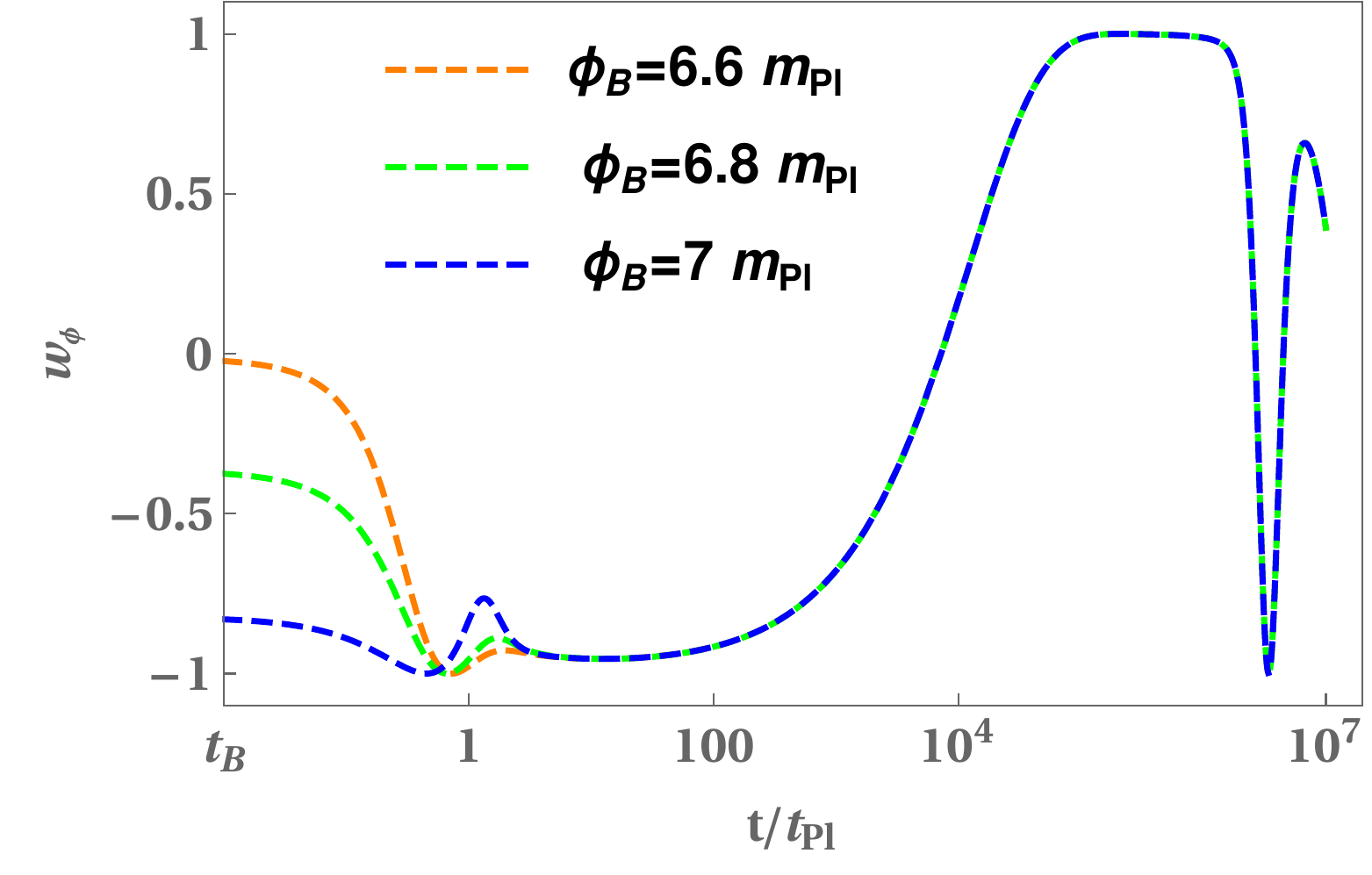}} & 
{\includegraphics[width=2.0in,height=1.6in,angle=0]{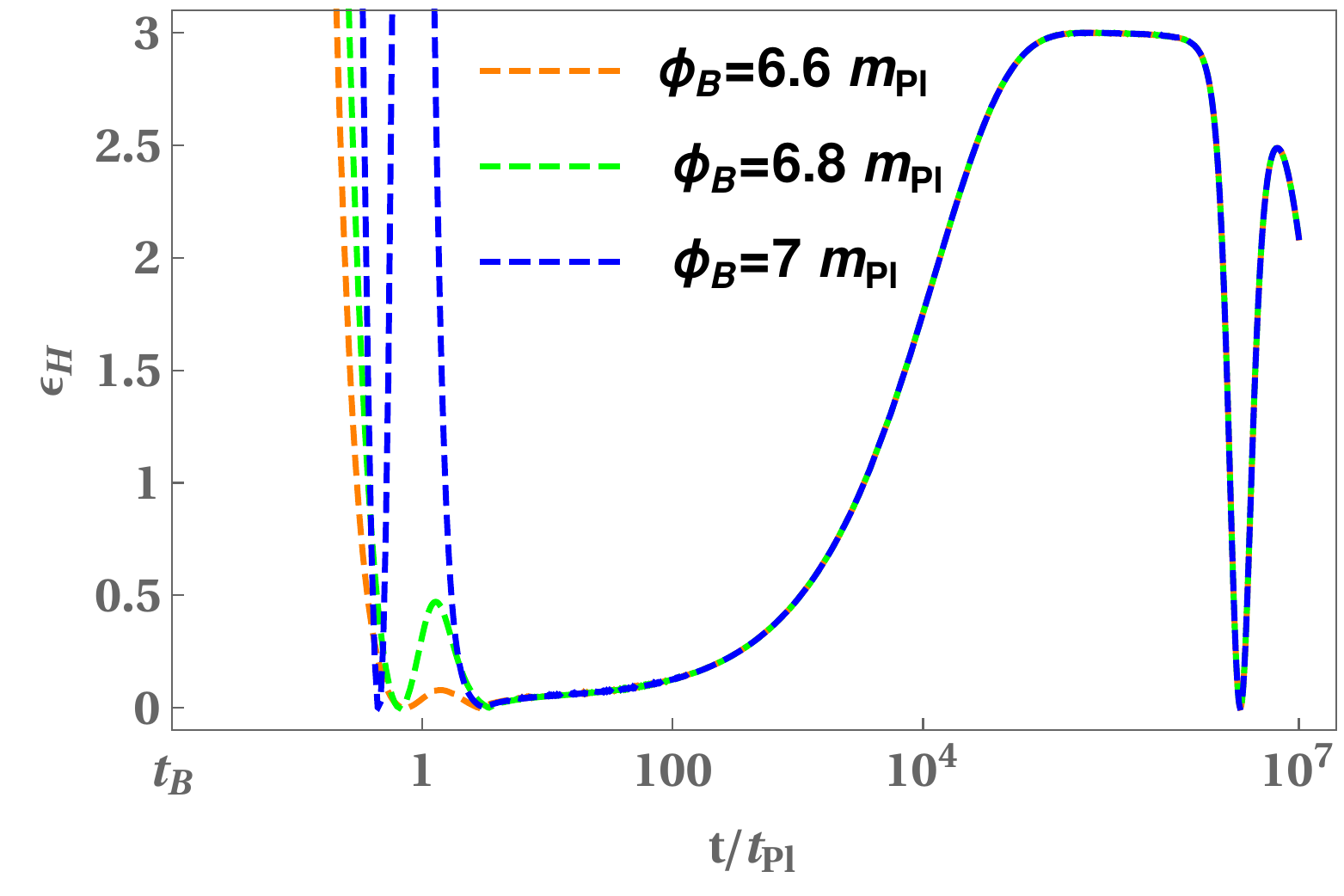}} 
\end{tabular}
\end{center}
\caption{ The figure displays the numerical results for $p=5, v=0.19 m_{Pl}$ and $ V_0=7.708 \times 10^{-17} m_{Pl}^4$. Similar to Fig. \ref{fig:P5vpt1} in this case also the slow-roll inflation is not obtained.}
\label{fig:P5v1}
\end{figure*}

Third, we evolve the background  equations (\ref{eq:Hub}) and (\ref{eq:ddphi}) with potential (\ref{eq:pot}) for $p=5$, $v=0.019 m_{Pl}$ and $V_0=3.577 \times 10^{-20} m_{Pl}^4$. The numerical results are illustrated in Fig. \ref{fig:P5vpt1}. Top panels correspond to KED case whereas bottom ones are for PED initial conditions. In both the cases, the numerical evolution of scale factor $a(t)$ is not exponential, the equation of state $w(\phi)$ and the slow-roll parameter $\epsilon_H$ provide oscillatory behavior which are not favored by the conditions of inflation. Furthermore, we examine the background  evolution for $p=5$, $v=0.19 m_{Pl}$ and $V_0=7.708 \times 10^{-17} m_{Pl}^4$. The numerical results are depicted in Fig. \ref{fig:P5v1}. By looking the evolution of $w(\phi)$ and $\epsilon_H$, the oscillatory behavior is reduced but still we don't obtain the desired slow-roll inflation.

Finally, we do brief discussion of the model under consideration. We display the shape of the potential (\ref{eq:pot}) as a function of $\phi$ in Fig. \ref{fig:pot}. For $p=2, 4$ and 6 (left panel), we observe that the potential has a plateau around $\phi  \approx 0$ and two minima at $\phi= \pm v$. The coefficient $p$ governs the flatness of the plateau and steepness of the potential. The larger $p$ corresponds to a long pronounced plateau and a more steepness around the minimum. According to Planck 2018 results \cite{Planck2018}, the Hilltop model with $p<4$ is not fit well with data. The predictions of the spectral index $n_s$ for $p=3$ and $N_{inf}=60$ is found to be 0.935484 which is in some tension with the most recent Planck bounds $n_s = 0.9649 \pm 0.0042$ (at 68 \% CL) \cite{Planck2018}. The predicted value of $n_s$ can be larger if $\phi^3/v^3$ in the potential is replaced by some higher power  $\phi^p/v^p$ with $p \geq 4$. The values of $n_s$ with $N_{inf}=60$ are found as 0.95122, 0.95652 and 0.95918 for $p=4, 5$ and 6, respectively.

\section{Phase Portrait}
\label{sec:port}
In this section, we shall display the phase space trajectories for the Hilltop potential (\ref{eq:pot}) with $\dot{\phi_B}>0$ and $\dot{\phi_B}<0$, and also for KED and PED initial conditions of inflaton field in the $(\phi/m_{Pl}, \dot{\phi}/m_{Pl}^2)$ plane. We choose the model parameters as  $p=4, v=2 m_{Pl}$ and $V_0=0.01 m_{Pl}^4$ for the better depiction of Fig. \ref{fig:port}. Since $\rho_c$ is the maximum energy density that constraints the initial value of $\phi_B$ such as $| \dot{\phi}_B |/m_{Pl}^2  < 0.91 $ and $\phi_B/m_{Pl} \in (-3.3, 3.3)$ under the chosen model parameters. Moreover, the initial data surface is totally finite due to the critical energy density
$\rho_c$ which is represented by the black boundary surface. In the Hilltop potential, the inflaton field rolls away from a local maximum at $\phi=0$ to minimum at $\phi= \pm v$. Therefore, we have two minima at $\phi=+v$ and $-v$ that are exhibited by red and blue color, respectively in Fig. \ref{fig:port}. All the trajectories onset from the surface of bounce ($\rho=\rho_c$) and move toward their respective minima, namely, $\phi=+v$ (red color) and $\phi=-v$ (blue color) which are the stable points. Regions close to the boundary belong to the higher energy density where the quantum effects are dominated whereas the lower energy density is found near the minima in $(\phi/m_{Pl}, \dot{\phi}/m_{Pl}^2)$ plane.

Let us compare our results with the power law and Starobinsky potentials. In the case of power law, both KED and PED initial values of inflaton field are consistent with observations in terms of number of $e$-folds \cite{alam2017} whereas Starobinsky potential is in good agreement with observations only for KED (except for small subset) initial conditions and not for PED ones \cite{Bonga2016}. In the case of Hilltop potential, we obtained the desired slow-roll inflation for both KED (except small subset) and PED initial conditions with $p=4$ and $v=1M_{Pl}$, and found physically viable initial values of inflaton field. However, for  other values of $p$ and $v$, the slow-roll inflation can not be achieved.

\section{Conclusions}
\label{sec:conc}
In this paper, we examined the dynamical behavior of pre-inflation for Hilltop potential (\ref{eq:pot}) in the framework of LQC. We found physical viable initial conditions of inflaton field at the bounce that produced the desired slow-roll inflation, and also generated enough number of $e$-folds. First, we studied the background equations (\ref{eq:Hub}) and (\ref{eq:ddphi}) for Hilltop potential (\ref{eq:pot}) with $p=4$, $v=0.019m_{Pl}$ and $V_0=2.632 \times 10^{-21} m_{Pl}^4$. The numerical results for KED (upper panels) and PED (lower panels) initial conditions of inflaton field at the bounce are presented in Fig. \ref{fig:P4vpt1}. We noticed that the evolution of $a(t)$ did not provide the exponential expansion which was required for inflation. Furthermore, the numerical evolution of $w(\phi)$ and $\epsilon_H$ exhibited oscillatory behavior. In the lower panels of Fig. \ref{fig:P4vpt1}, the numerical evolution of expansion factor is not consistent with the analytical solution (\ref{eq:a}). Also, $w(\phi)$ did not stay pegged at $-1$ and $\epsilon_H$ is not less than unity. Hence, in this case, the slow-roll inflation can not be achieved.

Second, we evolved the potential (\ref{eq:pot}) with the background equations (\ref{eq:Hub}) and (\ref{eq:ddphi}) for $p=4$, $v=0.19m_{Pl}$ and $V_0=2.632 \times 10^{-17} m_{Pl}^4$. The results are displayed in Fig. \ref{fig:P4v1}. In case of KED initial conditions (upper panels), the behavior of expansion factor $a(t)$ is universal in the bouncing regime, and well approximated by analytical solution (\ref{eq:a}). Later, it showed exponential expansion. The contribution of the potential in the bouncing phase was very small as compared to the kinetic one. Therefore, it showed almost negligible effects on the evolution of background that gave rise to universal behavior of expansion factor. The evolution of $w(\phi)$ (upper middle panel) demonstrated that the background evolution before preheating was divided into three different phases: {\it bouncing, transition and slow-roll inflation}. In the bouncing regime, $w(\phi) \simeq +1$, it decreases to $-1$ during the transition regime whereas in the slow-roll phase $w(\phi)$ is close to $-1$ untill the end of slow-roll inflation. By looking the evolution of $\epsilon_H$ (upper right panel), in the bouncing phase $\epsilon_H > 1$ which reduces to $\epsilon_H \approx 0$ during the transition phase, and remains so untill the end of slow-roll inflation. However, the whole range of $\phi_B$ did not provide the inflationary phase. In other words, a subset of KED initial values of inflaton field was also present that provide non-inflationary regime, see middle panels of Fig. \ref{fig:P4v1} and Table \ref{tab-phiB}. In case of the PED initial conditions (lower panels), the universality of $a(t)$ disappeared, bouncing and transition phases no longer existed. Though, slow-roll inflation can still be obtained. We also found the number of $e$-folds for the desired slow-roll inflation. For successful inflation at least 60 $e$-folds are required to be compatible with observation. We displayed important inflationary parameters in Table \ref{tab} for different values of $\phi_B$. From the table, one can inferred that the number of $e$-folds increased as $\phi_B$ grew. Next, we evolved the potential (\ref{eq:pot}) with equations (\ref{eq:Hub}) and (\ref{eq:ddphi}) for $p=5$, $v=0.019m_{Pl}$, $V_0=3.577 \times 10^{-20} m_{Pl}^4$ and
$p=5$, $v=0.19m_{Pl}$ and $V_0=7.708 \times 10^{-17} m_{Pl}^4$. The results are exhibited in Figs. \ref{fig:P5vpt1} and \ref{fig:P5v1}, respectively. By looking both the figures, we concluded that neither KED nor PED initial conditions of inflaton field at the bounce generated the desired slow-roll inflation as the behavior of evolution of $w(\phi)$ and $\epsilon_H$ found to be oscillatory.

\begin{figure*}[tbp]
\begin{center}
\begin{tabular}{c}
{\includegraphics[width=2.5in,height=2in,angle=0]{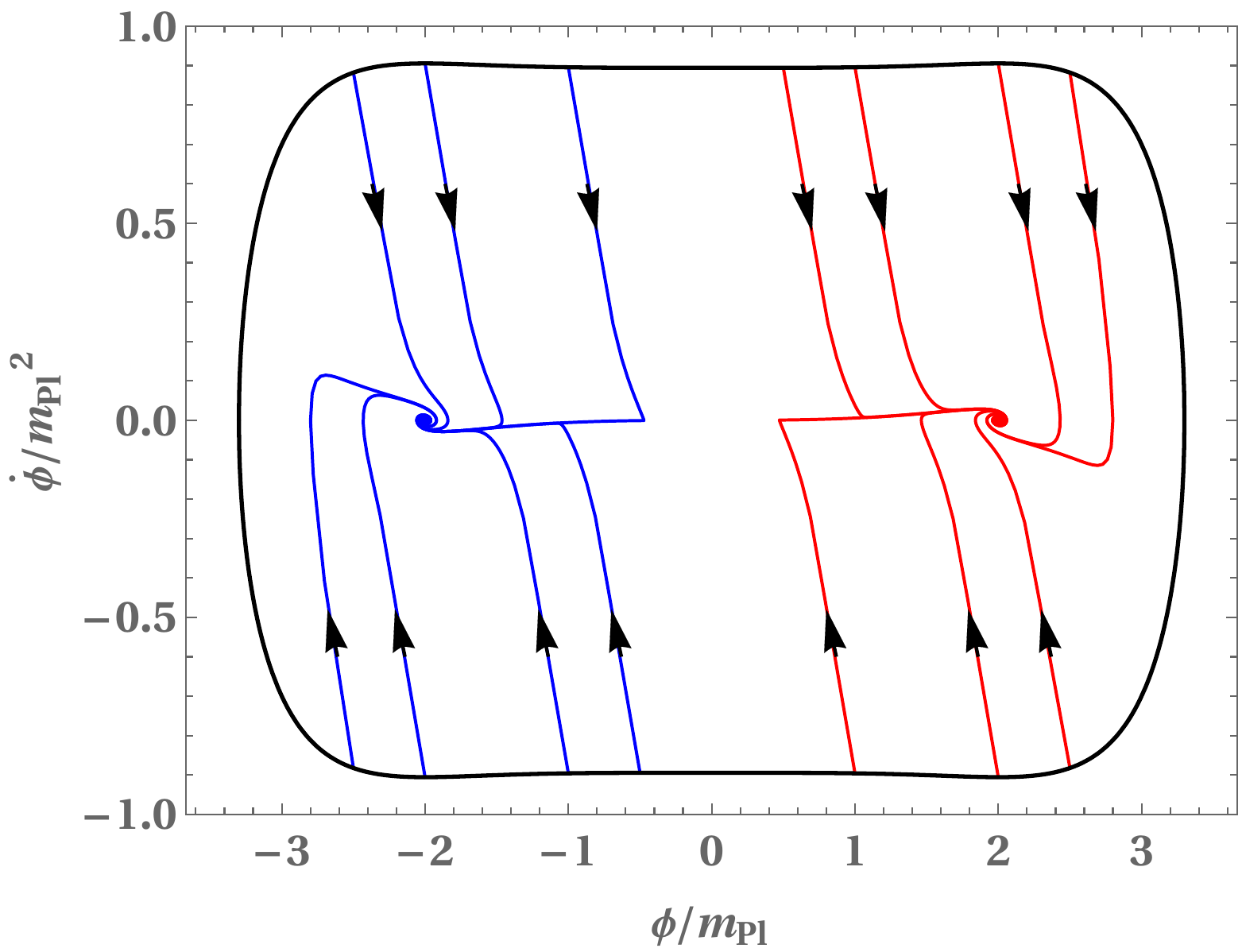}} 
\end{tabular}
\end{center}
\caption{This figure exhibits the phase portrait for Hilltop potential (\ref{eq:pot}) with both PIV and NIV in the $(\phi/m_{Pl}, \dot{\phi}/m_{Pl}^2)$ plane. All trajectories with arrowheads onset from the bounce where $\rho=\rho_c$ (boundary surface), and end at minimum value of $\phi$, namely, $\phi=+v$ (red) and $-v$ (blue) which are the attractor points. For better depiction, we choose $p=4, v=2 m_{Pl}$ and $V_0=0.01 m_{Pl}^4$. }
\label{fig:port}
\end{figure*}

Finally, we depicted the phase portrait for Hilltop potential (\ref{eq:pot}) with PIV and NIV, and also for KED and PED initial conditions of inflaton field in $(\phi/m_{Pl}, \dot{\phi}/m_{Pl}^2)$ plane, see Fig. \ref{fig:port}. For better depiction, we used $p=4, v=2 m_{Pl}, V_0=0.01 m_{Pl}^4$. The critical energy density $\rho_c$ restricts the values of $\phi_B$ as the Hilltop potential is unbounded from the above. Therefore, we got the compact surface at the bounce due to $\rho_c$. The finite data surface is denoted by the black boundary curve in Fig. \ref{fig:port} where $| \dot{\phi}_B |/m_{Pl}^2  < 0.91 $ and $\phi_B/m_{Pl} = \pm 3.3$ under the chosen parameters. The potential (\ref{eq:pot}) has a local maximum at $\phi=0$ and two minimum at $\phi= \pm v$. Therefore, the trajectories start from the bounce  ($\rho=\rho_c$) and directed toward their respective minimum, namely, $\phi=+v$ (red) and $-v$ (blue) which are the stable points and behaves as the attractor. The quantum geometric effects are dominated at the quantum bounce. The regions near the bounce have maximum energy density and near the minima, lower density is found in $(\phi/m_{Pl}, \dot{\phi}/m_{Pl}^2)$ plane.

\section*{Acknowledgments}
The work is partially financially supported by the Ministry of Education and Science
of the Republic of Kazakhstan, Grant No. AP08856912.

\section*{Data Availability}
Data will be made available on reasonable request.



\end{document}